\documentclass[iop]{emulateapj}	

\submitted{Received 2011 November 10; accepted 2012 February 8}

\shorttitle{THE FORMATION OF THE MILKY WAY THROUGH COLD HALO SUBSTRUCTURE. III.}
\shortauthors{SCHLAUFMAN ET AL.}

\begin{document}

\title{INSIGHT INTO THE FORMATION OF THE MILKY WAY THROUGH COLD HALO
SUBSTRUCTURE. III. STATISTICAL CHEMICAL TAGGING IN THE SMOOTH HALO}

\author{Kevin C. Schlaufman\altaffilmark{1},
Constance M. Rockosi\altaffilmark{1,2,8}, Young Sun Lee\altaffilmark{3},
Timothy C. Beers\altaffilmark{3,4}, Carlos Allende Prieto\altaffilmark{5,6},
Valery Rashkov\altaffilmark{1}, Piero Madau\altaffilmark{1},
and Dmitry Bizyaev\altaffilmark{7}}

\altaffiltext{1}{Astronomy and Astrophysics Department, University
of California, Santa Cruz, CA 95064, USA; kcs@ucolick.org,
crockosi@ucolick.org, valery@ucolick.org, pmadau@ucolick.org}
\altaffiltext{2}{University of California Observatories, Santa Cruz,
CA 95064, USA}
\altaffiltext{3}{Department of Physics and Astronomy and JINA: Joint Institute
for Nuclear Astrophysics, Michigan State University, E. Lansing, MI 48824,
USA; lee@pa.msu.edu, beers@pa.msu.edu}
\altaffiltext{4}{National Optical Astronomy Observatories, Tucson,
AZ 85719, USA}
\altaffiltext{5}{Instituto de Astrof\'{\i}sica de Canarias, 38205 La
Laguna, Tenerife, Spain; callende@iac.es}
\altaffiltext{6}{Departamento de Astrof\'{\i}sica, Universidad de La
Laguna, 38206 La Laguna, Tenerife, Spain}
\altaffiltext{7}{Apache Point Observatory, P.O. Box 59, Sunspot, NM 88349,
USA; dmbiz@apo.nmsu.edu}
\altaffiltext{8}{Packard Fellow}

\begin{abstract}

\noindent
We find that the relative contribution of satellite galaxies accreted at
high redshift to the stellar population of the Milky Way's smooth halo
increases with distance, becoming observable relative to the classical
smooth halo about 15 kpc from the Galactic center.  In particular,
we determine line-of-sight-averaged [Fe/H] and [$\alpha$/Fe] in the
metal-poor main-sequence turnoff (MPMSTO) population along every Sloan
Extension for Galactic Understanding and Exploration (SEGUE) spectroscopic
line of sight.  Restricting our sample to those lines of sight along which
we do not detect elements of cold halo substructure (ECHOS), we compile
the largest spectroscopic sample of stars in the smooth component of the
halo ever observed in situ beyond 10 kpc.  We find significant spatial
autocorrelation in [Fe/H] in the MPMSTO population in the distant half of
our sample beyond about 15 kpc from the Galactic center.  Inside of 15
kpc however, we find no significant spatial autocorrelation in [Fe/H].
At the same time, we perform SEGUE-like observations of $N$-body
simulations of Milky Way analog formation.  While we find that halos
formed entirely by accreted satellite galaxies provide a poor match
to our observations of the halo within 15 kpc of the Galactic center,
we do observe spatial autocorrelation in [Fe/H] in the simulations at
larger distances.  This observation is an example of statistical chemical
tagging and indicates that spatial autocorrelation in metallicity is a
generic feature of stellar halos formed from accreted satellite galaxies.

\end{abstract}


\section{Introduction}

\defcitealias{schl09}{S09}
\defcitealias{schl11}{S11}

The smooth (or classical) halo of the Milky Way is the subset of the
larger halo population that is not in identified substructure that stands
out in some way from the local background, either photometrically,
kinematically, or chemically.  These halo substructures have been
discovered in star counts, kinematic measurements, and chemical abundances
\citep[e.g.,][]{tot98,tot00,ive00,yan00,ode01,viv01,gil02,new02,roc02,maj03,yan03,roc04,duf06,belo06a,belo06b,gri06a,gri06b,viv06,belo07,ses07,bell08,jur08,gri09,wat09,dej10,maj96,chi98,hel99,chi00,kep07,ive08,kle08,sea08,kle09,schl09,smi09,sta09,ive08,an09,schl11,caro12}.
Despite the large number of substructure detections, the majority of
the stellar content of the halo within 20 kpc of the Galactic center
is both expected theoretically \citep[e.g.,][]{mac11} and observed
\citep[e.g.,][]{bell08,schl09} to be in a kinematically smooth population.

While observations of the halo stellar population suggest that the
fraction of the halo in substructure increases with Galactocentric radius
\citep[e.g.,][]{bell08,schl09,xue11}, the fractions of the smooth halo
stellar population formed in situ, accreted in the major mergers that
formed the bulk of the halo, or accreted in subsequent satellite galaxy
minor mergers are observationally unconstrained.  The ratio of in situ
star formation to accretion and its dependence on distance from the
Galactic center may be related to the accretion history of Milky Way
analog halos \citep[e.g.,][]{zol09,zol10,hel11,fon11}, and the expectation
from simulations is that the contribution of the accreted component
likely increases with distance \cite[e.g.,][]{zol09,zol10,fon11}.

The stellar population in the classical smooth halo seems to
be kinematically smooth on large scales and has three other main
properties: (1) it is old, (2) its [Fe/H] distribution peaks at [Fe/H]
$\sim -1.6$ with full width at half maximum $\gtrsim 1$ dex, and (3) it
is enhanced in the $\alpha$-elements O, Mg, Si, Ca, and Ti relative to
iron \citep[e.g.,][]{rya91a,rya91b,mcw95,all06,schl09}.  \citet{rob05}
explained the age, metallicity, $\alpha$-enhancement, and kinematic
structure of the smooth component of the halo in the context of the
$\Lambda$CDM model of galaxy formation with the accretion of massive
$M_{\mathrm{tot}} \sim 5 \times 10^{10}~M_{\odot}$ halos $\sim\!\!10$
Gyr in the past.  The high mass and short timescale for star formation
in such massive progenitors of the smooth halo are consistent with the
observed chemistry.  In a suite of six simulations of Milky Way analog
halos, \citet{coo10} also found that halos acquire the bulk of their
mass from fewer than five significant contributors.  The \citet{rob05}
and \citet{coo10} scenario naturally explains the low [Fe/H] and high
[$\alpha$/Fe] in the halo.  On the other hand, the halo abundance pattern
is in contrast to the composition of the (surviving) classical dwarf
spheroidal (dSph) galaxies, which at the average [Fe/H] of the smooth halo
have [$\alpha$/Fe] closer to solar \citep[e.g.,][]{mat98,kir10,kir11a}.

In the first paper in this series \citep[][S09 hereafter]{schl09}, we
described the results of a systematic, statistical search for elements of
kinematically-cold halo substructure (ECHOS) in the inner halo metal-poor
main-sequence turnoff (MPMSTO) population.  One by-product of the search
for ECHOS detailed in \citetalias{schl09} is a catalog of MPMSTO stars
more than 4 kpc from the Galactic plane, more than 10 kpc from the
center of the Galaxy, within 17.5 kpc of the Sun, and free of both
surface-brightness and radial-velocity substructure.  We subsequently
refer to this sample as the pure smooth halo sample.  Though there is
likely still very diffuse substructure in the catalog, it has been cleaned
of substructure to the greatest extent possible using existing data.
In the second paper in this series \citep[][S11 hereafter]{schl11},
we analyzed co-added MPMSTO spectra to derive the average [Fe/H] and
[$\alpha$/Fe] for ECHOS and for the smooth component of the halo along
the same line of sight as each ECHOS.  We found that the MPMSTO stars
in ECHOS were systematically more metal rich and less [$\alpha$/Fe]
enhanced than the MPMSTO stars in the smooth component of the halo.
We concluded that the chemical abundance pattern of ECHOS was best matched
by a massive dSph galaxy with $M_{\mathrm{tot}} \gtrsim 10^{9}~M_{\odot}$.

In this third paper of the series, we quantify the degree of spatial
chemical inhomogeneity and spatial variation in chemical abundance in the
smooth component of the halo, using the substructure-cleaned sample of
MPMSTO stars produced in \citetalias{schl09} and the chemical abundance
technique developed in \citetalias{schl11}.  We also investigate the
dependence of these properties on distance from the Galactic center.

This concept is analogous to the idea of chemical tagging employed in the
solar neighborhood.  Stars formed within the same star forming region have
very similar abundances \citep[e.g.,][]{des06,des07a,des07b,bub10,pom11},
and this similarity has led to the use of chemical tagging in
concert with kinematic information to identify groups of stars
in the solar neighborhood with a common origin.  The chemical
homogeneity of stars formed in the same star forming region may be
generic \citep[e.g.,][]{bla10}.  Likewise, many halo substructures are
chemically distinct from the kinematically smooth halo stellar population
\citep[e.g.,][]{ive08,an09,schl11}.

To illustrate further, consider the following thought experiment.
Imagine that a Milky Way analog halo can be formed entirely in one of
two ways: (1) a scenario in which the smooth stellar halo formed in
a few dissipative major mergers (in which violent relaxation may also
have been important) and (2) a scenario in which the smooth stellar halo
formed by the accretion and tidal disruption of satellite galaxies (in
which diffusion processes are negligible).  These two limiting cases are
vastly simplified and cannot accurately reproduce the observed properties
of the Milky Way.  Nevertheless, they are illustrative of two opposite,
extreme models of halo formation.

In the stellar halo formed entirely in a few dissipative major mergers,
the resultant stellar population should be very well mixed.  Consequently,
an observer measuring metallicity along many lines of sight closely spaced
on the sky and in the same distance range throughout the halo would find
that lines of sight that are closely spaced are no more likely to have
similar chemical abundances than lines of sight that are widely separated.

In the stellar halo formed entirely by the accretion and tidal disruption
of satellite galaxies, an observer measuring metallicity along many lines
of sight and in the same distance range throughout the halo would find
that lines of sight that are close together on the sky are also likely
to be close together in metallicity.  This effect arises simply because
lines of sight that are close together on the sky in a small range
of distance are more likely to sample the stellar debris of only a few
recently disrupted satellite galaxies.  On the other hand, lines of sight
that are widely separated on the sky would be very unlikely to sample
the debris of the same accreted galaxy.  As a result of this effect,
the observer would see spatially-correlated metallicities in the halo.

We measure the degree of spatial chemical inhomogeneity and spatial
variation in chemical abundance in the smooth component of the halo
to differentiate between these two possibilities as a function of
distance from the Galactic center.  This paper is organized as follows:
in Section 2 we define the data we use in this analysis.  In Section 3,
we describe how we determine line-of-sight average metallicities in the
kinematically smooth halo.  We also report how we quantify the degree
of spatial autocorrelation in the halo of the Milky Way and in halos of
simulated Milky Way analogs.  In Section 4, we discuss the implications
of our findings for the formation of the Milky Way.  We summarize our
conclusions in Section 5.

\section{Data}

The Sloan Extension for Galactic Understanding and Exploration
(SEGUE) survey observed approximately 240,000 Milky Way stars with
apparent magnitudes in the range $14 < g < 20.3$ with the fiber-fed
Sloan Digital Sky Survey (SDSS) spectrograph at moderate resolution.
Spectroscopic targets were selected from the combined 11,663 deg$^2$
$ugriz$ photometric footprint of the SDSS and SEGUE.  The SDSS telescope
and spectrograph obtain $R \approx 1800$ spectra between 3800 \AA~and
9000 \AA.  The SEGUE data processing pipelines, survey strategy, and
radial velocity and atmospheric parameter accuracies are described in
\citet{yan09}, \citet{lee08a,lee08b,lee11}, \citet{all08}, \citet{smo11},
and the SDSS-II DR7 paper \citep{abaz09}.  The SDSS survey and its
instrumentation are described in detail in \citet{fuk96}, \citet{gun98},
\citet{yor00}, \citet{hog01}, \citet{smi02}, \citet{pie03}, \citet{ive04},
\citet{gun06}, \citet{tuc06}, and \citet{pad08}.

In \citetalias{schl09}, we described the results of a systematic,
statistical search for ECHOS in the inner halo MPMSTO population.
We defined the inner halo as the volume more than 10 kpc from the
Galactic center, within 17.5 kpc of the Sun, and more than 4 kpc from the
Galactic plane.  While we did find statistically significant evidence of
substructure along 19 of 137 7 deg$^2$ lines of sight that we searched,
on average we found no reason to reject a kinematically smooth model
for the MPMSTO population in that volume.

In this paper, we examine the MPMSTO population along those 118 7
deg$^2$ lines of sight where there is no significant surface-brightness
substructure and for which we found no significant radial-velocity
substructure in \citetalias{schl09}.  This results in a sample of
9,005 MPMSTO stars in the same volume with both photometric and
spectroscopic data.  These MPMSTO stars have both the $g-r$ color
and significant UV excess expected for the main-sequence turnoff of a
metal-poor population (for a detailed description of the MPMSTO sample
see Section 2 of \citetalias{schl09}).  Given the magnitude limits of
the SEGUE survey, the MPMSTO sample was selected because MPMSTO stars
are the highest-density tracer of the inner halo.  At the mean $g-r$
color and metallicity of our sample, SEGUE radial velocities are precise
to 5 km s$^{-1}$ at $g=18$ and 20 km s$^{-1}$ at $g=20.3$.

\begin{figure}
\plotone{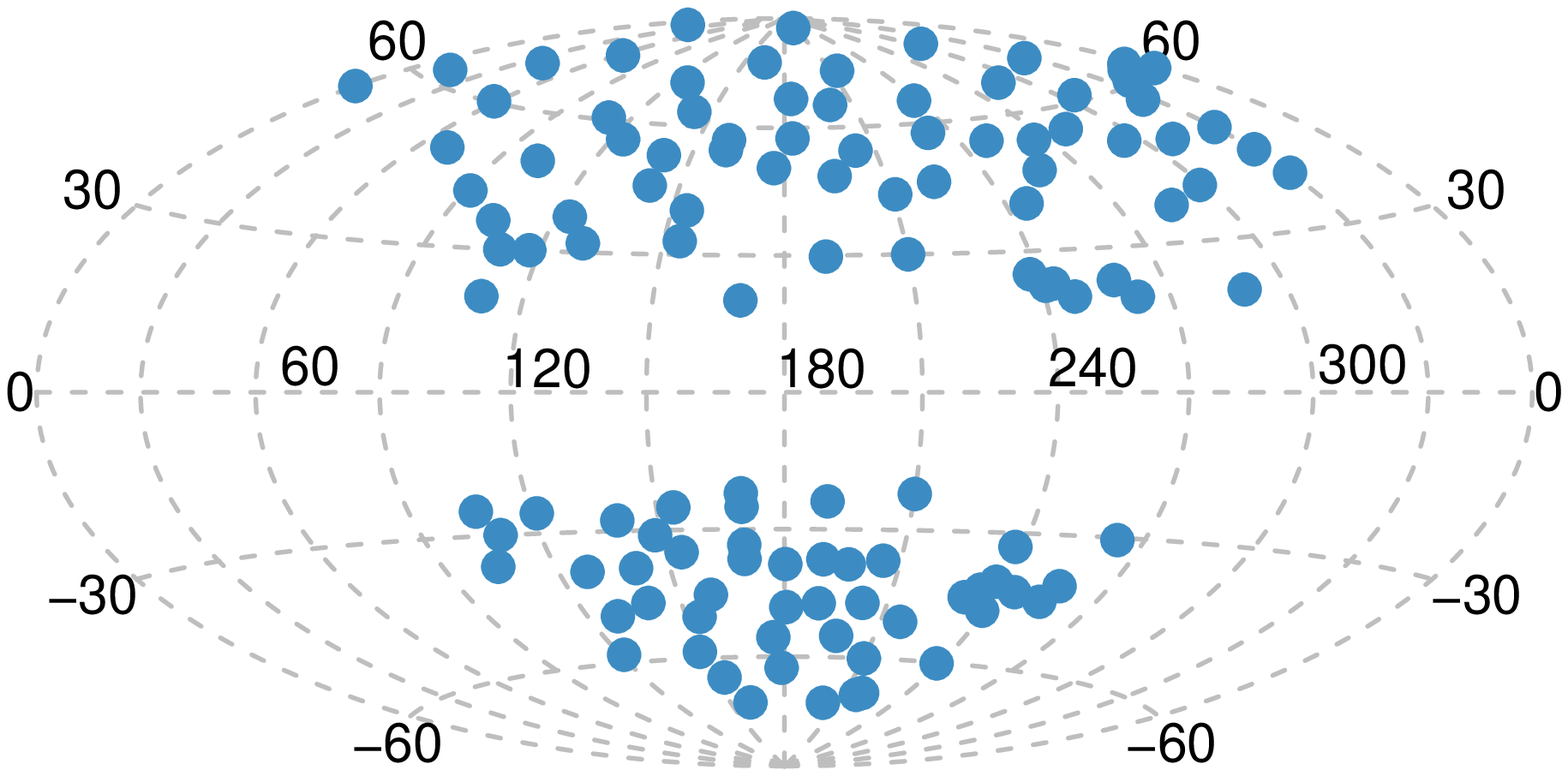}
\caption{ECHO-free pure smooth halo sample line-of-sight spatial
distribution in Galactic coordinates.\label{fig01}}
\end{figure}

This sample is the largest spectroscopic sample of halo stars in
the smooth component beyond 10 Galactocentric kpc ever assembled.
To the extent possible with current data, it has been cleaned of
both surface-brightness and radial-velocity substructure.  As such,
it allows for an unprecedented analysis of the chemical properties of
the classical smooth halo.  We plot in Figure~\ref{fig01} the spatial
distribution of lines of sight populated by MPMSTO stars in the pure
smooth halo population.

\section{Analysis}

For each SEGUE line of sight in our pure smooth halo sample described in
Section 2, we compute the average [Fe/H]\footnote{[Fe/H] is the context of
this paper is the total metallicity assuming all metals are coupled, not
truly the iron abundance.} and [$\alpha$/Fe] of the MPMSTO population
along that line of sight as a function of distance from the Galactic
center.  We subdivide our sample into nearby and distant subsamples
and compare the average [Fe/H] and [$\alpha$/Fe] in these subsamples
to investigate the effect of distance from the Galactic center on the
chemical properties of the smooth halo.  We quantify the inhomogeneity
and spatial variation in average chemical abundances by measuring the
spatial autocorrelation of the line-of-sight-averaged chemical abundances.
To better understand our observed result for the Milky Way, we perform
SEGUE-like observations of both the \citet{bul05} suite of halos and the
Via Lactea 2 halo \citep[e.g.,][]{die08,mad08,zem09,ras11} to determine
the degree of spatial chemical inhomogeneity and spatial variation in
chemical abundance present in $N$-body simulations of Milky Way analog
halos.

\subsection{Line of Sight Averaged Chemical Abundances}

For each MPMSTO star in the pure smooth halo sample described in Section
2, we use the M~13 fiducial sequence from \citet{an08} to very roughly
estimate its distance from the center of the Galaxy based on its $g-r$
color and apparent $r$-band magnitude (assuming that the Sun is 8 kpc
from the center of the Galaxy).

In order to quantify the spatial variation in chemical abundance in the
smooth halo (and their dependence on distance from the Galactic center),
we subdivide our sample into nearby and distant subsamples.  The halo
stellar population may change gradually with distance, and our division
into nearby and distant regions can only crudely capture such a change.
However, by splitting the halo coarsely into nearby and distant regions,
we can compute an average value for [Fe/H] and [$\alpha$/Fe] in two
volumes along each line of sight.  We can then vary the distance at which
the larger sample is split into the nearby and distant subsample in order
to explore how the average stellar population varies from line of sight
to line of sight at constant distance from the Galactic center, and how
the amplitude and spatial scale of that variation change with distance.
We use three different distance thresholds to divide our sample into
nearby and distant regions: 12.1, 14.1, and 16.8 kpc.  These are the
quartiles of the estimated Galactocentric distance distribution of
our sample.  For each line of sight, the result is six overlapping
subsamples in distance from the Galactic center.

We use the spectral co-addition technique summarized in Appendix A
and described in detail in \citetalias{schl11}.  For each subsample
and line of sight, we create the ``average" MPMSTO spectrum for that
subsample and line of sight by creating a single co-added spectrum from
the MPMSTO spectra in that subsample and line of sight.  Co-addition
is necessary in this situation because the apparent faintness of the
MPMSTO stars in our distant subsamples results in spectral S/N per pixel
in individual spectra that are not high enough to obtain precise and
accurate stellar parameters from the SEGUE Stellar Parameter Pipeline
\citep[SSPP;][]{lee08a,lee08b,lee11,all08,smo11}.  We use the [Fe/H] and
[$\alpha$/Fe] reported by the SSPP.  In both cases, the Fe abundance
calculated is an overall metallicity, not just the abundance of iron.

We verified the accuracy and precision of the co-addition procedure
using both MPMSTO stars observed by SEGUE in the globular clusters
M~13 and M~15 and SEGUE observations of field MPMSTO stars with stellar
parameters derived from subsequent high-resolution spectroscopy.  The mean
square error (MSE $\equiv$ bias$^2$ + variance) of our co-added abundance
analysis ranges from 0.1 dex in [Fe/H] at [Fe/H] $\approx -1.0$ to 0.20
dex in [Fe/H] for [Fe/H] $\approx -1.5$.  Our MSE estimate includes
the effect of systematic error, which will not affect the spatial
autocorrelation analysis we describe in this paper.  For that reason,
these error estimates are conservative upper limits on our errors.

\begin{figure}
\epsscale{0.8} 
\plotone{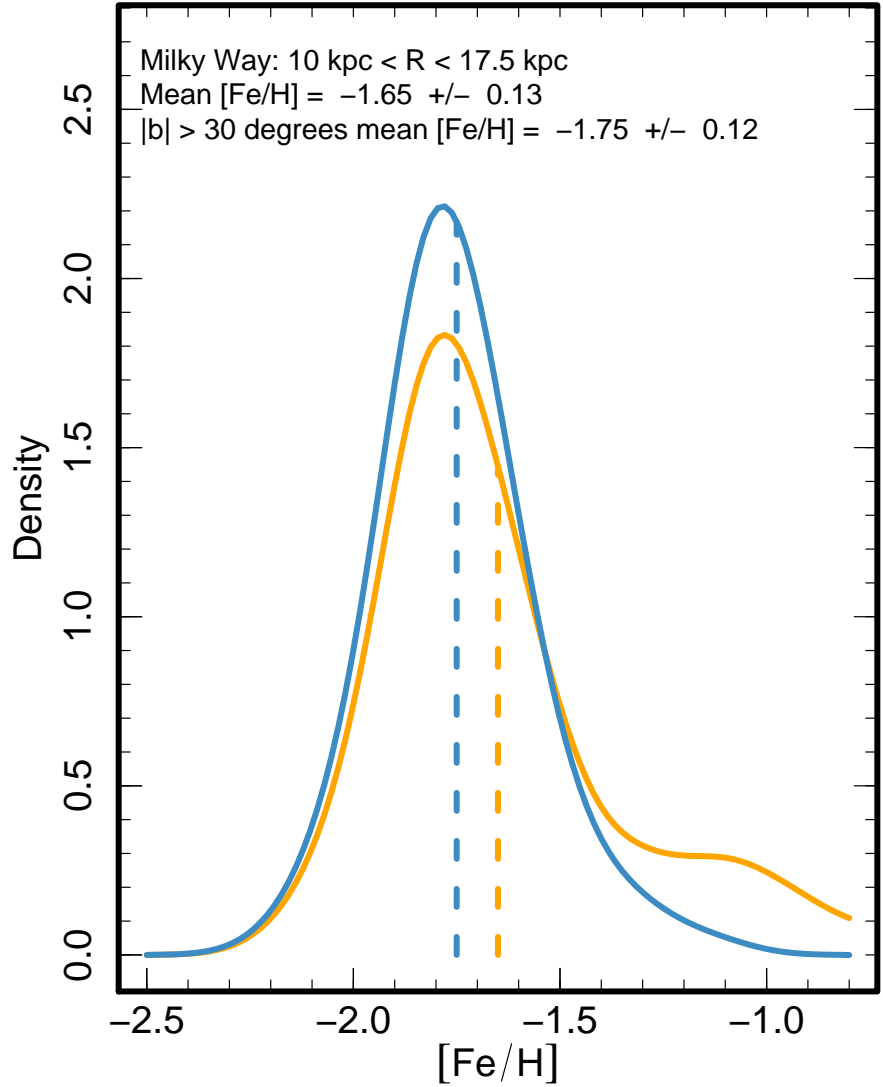}
\caption{Density of line-of-sight-averaged metallicity distribution
between 10 and 17.5 kpc from the Galactic center in the smooth
component of the halo kernel smoothed to 0.1 dex.  The orange curve
is the distribution for all lines of sight, while the blue curve
is the distribution for lines of sight with $|b| > 30^{\circ}$.
The mean [Fe/H] of each subsample is indicated by the vertical
dashed line.  Our estimates of the mean metallicity include both
random ($\sim\!\!0.01$ dex) and systematic error ($\sim\!\!0.1$
dex).  The distribution for the full sample has a heavy tail to high
metallicity that is not present at high Galactic latitude.  As a result,
the mean of the high Galactic latitude sample is systematically more
metal poor by about 0.1 dex.  This offset is likely related to the
metal-rich low-latitude substructure first identified in Monoceros
\citep[e.g.,][]{new02,belo06b,jur08,ive08,schl11}.\label{fig02}}
\end{figure}

\begin{figure}
\plotone{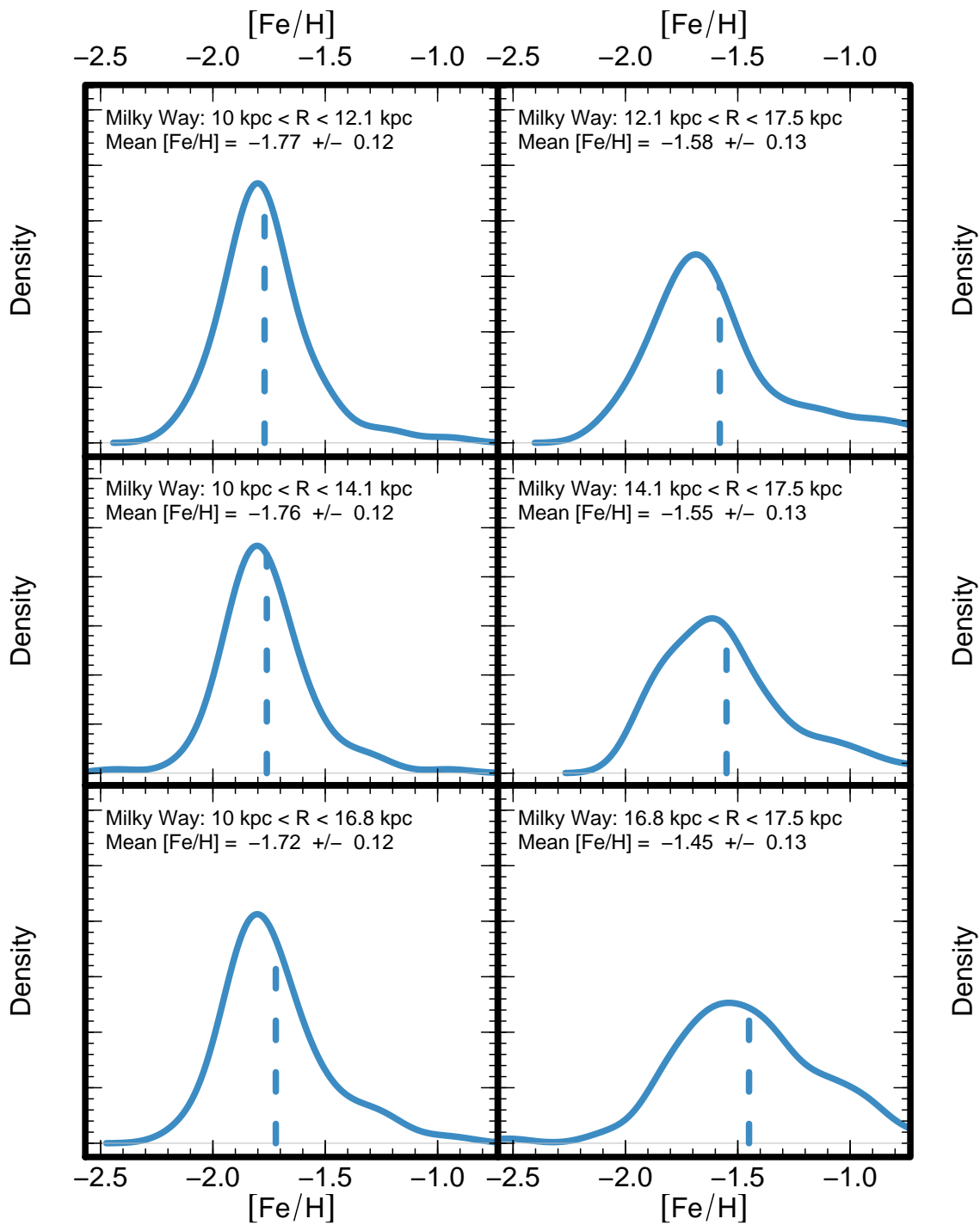}
\caption{Density of line-of-sight-averaged metallicity distribution as
a function of distance from the Galactic center in the smooth component
of the halo kernel smoothed to 0.1 dex.  The $y$-axis scale is the same
for each panel, and its numerical range is arbitrary such that the area
under each curve integrates to 1.  We split our tracer sample in two
very roughly in distance at three different split points: 12.1, 14.1 and
16.8 kpc.  Each row corresponds to the result for a single split-point.
In all cases, the nearby part of the sample is in the left column and
the distant part of the sample is in the right column.  The mean [Fe/H]
of each subsample is indicated by the vertical dashed line.  Our mean
metallicities are in the range  $-1.8 \lesssim$ [Fe/H] $\lesssim -1.5$.
We give the average [Fe/H] and uncertainties in Table~\ref{tbl-3}.
We observe no significant metallicity gradient in our sample, though
there may be a hint of a positive metallicity gradient at large
radius.\label{fig03}}
\end{figure}

We give the result of these calculations in Tables~\ref{tbl-1}
and~\ref{tbl-2}.  We plot the line-of-sight-averaged [Fe/H] distribution
from Table~\ref{tbl-1} in Figures~\ref{fig02} and \ref{fig03}.
We plot the distribution of the data in Tables~\ref{tbl-1} and
\ref{tbl-2} in the [Fe/H]--[$\alpha$/Fe] plane in Figure~\ref{fig04}.
Close to the Galactic center, the halo is dominated by the classical
$-1.8 \lesssim$ [Fe/H] $\lesssim -1.5$ and [$\alpha$/Fe]-enhanced
population \citep[e.g.,][]{rya91a,rya91b,mcw95,all06}.  The spread
in [$\alpha$/Fe] is consistent with a homogeneous [$\alpha$/Fe]
$\approx 0.3$ given our errors.  We find $-1.8 \lesssim$ [Fe/H]
$\lesssim -1.5$ and [$\alpha$/Fe] $\approx 0.3$ in the halo
MPMSTO population at 10 kpc.  These values are consistent with
high-resolution abundance analyses of local halo stars at 100 pc
\citep[e.g.,][]{edv93,nis97,han98,ful00,pro00,ste02,ben03,red03,ven04}.

Farther from the Galactic center, the halo has a non-negligible
dSph-like component with [Fe/H] $\approx -1.0$ and [$\alpha$/Fe]
$\approx 0.1$ \citep[e.g.,][]{kir10,kir11a,kir11b}.  This component
is found preferentially at $|b| \lesssim 30^{\circ}$, indicating a
possible association with the low-latitude substructure first identified
in Monoceros \citep[e.g.,][]{new02,belo06b,jur08,ive08,schl11} and
perhaps with the disk of the Milky Way.  The chemical abundances we
find in the smooth halo at low Galactic latitude are very similar to
the chemical abundances we derived in \citetalias{schl11} for this
low-latitude substructure.  This low-latitude substructure is not
apparent in [$\alpha$/Fe] alone, as the precision of our measurements
is much worse in [$\alpha$/Fe] than for [Fe/H].  The observation of
this chemical abundance pattern exclusively at low Galactic latitude
indicates that it may be related to satellite-induced disk flaring
\citep[e.g.,][]{kaz08,kaz09,pur10,pur11}.

Table~\ref{tbl-3} provides the mean [Fe/H] of the smooth halo in each
Galactocentric distance slice, both in the entire sample and only for
lines of sight with $|b| > 30^{\circ}$.  From both our current data
and our analysis in \citetalias{schl11}, we have reason to believe
that the known substructure at low Galactic latitude is metal rich
and quite prominent at the largest Galactocentric distances probed by
our MPMSTO sample.  For that reason, we focus on the mean [Fe/H] above
$|b| = 30^{\circ}$.  In that case, there are no instances in which mean
[Fe/H] inferred for the inner subsample differs significantly from the
mean [Fe/H] inferred for its complement outer subsample.  The largest
difference we measure is for the volume $10 < d < 16.8$ kpc and its
complement volume $16.8 < d < 17.5$ kpc: [Fe/H] $= -1.77 \pm 0.12$ versus
[Fe/H] $= -1.53 \pm 0.13$.  That difference is only significant at the
1.4$\sigma$ level.  As such, it can only be regarded as a very tentative
hint that there may be a positive metallicity gradient in the halo.
As a result, we find no significant [Fe/H] gradient in the volume probed
by our MPMSTO sample.

We have estimated the metallicity bias of our UV excess selection for the
MPMSTO sample.  Relative to stars with [Fe/H] $= -1.6$, stars at [Fe/H]
$= -3$ are twice as likely to be selected for spectroscopic follow-up,
while stars at [Fe/H] $= -1.1$ are half as likely to be selected.  The
selection probability follows a smooth, approximately parabolic relation
over that range in metallicity, and cuts off rapidly below [Fe/H] $= -1$.
Modeling this selection bias and including our measurement errors, for a
halo population with true mean [Fe/H] $= -1.6$ and standard deviation 0.4
\citep[e.g.,][]{rya91a,rya91b} we expect to measure a mean value of [Fe/H]
$= -1.79$, in good agreement with the means in Table~\ref{tbl-3}.  All of
the metallicity distributions in Figure~\ref{fig03} are consistent with
a majority population with mean near [Fe/H] $= -1.6$, in agreement with
the measurements of \citet{rya91a,rya91b} plus an additional population
of more metal-rich stars.  The selection bias in our sample is too mild
for the true population means to be significantly more metal rich than
about [Fe/H] $= -1.5$.

\subsection{Simulated Observation of Theoretical Models}

\subsubsection{Theoretical Models Used}

We simulate SEGUE-like observations of the \citet{bul05} suite of halos
as well as the Via Lactea 2 halo.  The \citet{bul05} simulations used a
``hybrid" approach in which they modeled accretion events onto a Milky
Way-like galaxy with $N$-body particles while modeling the rest of the
galaxy with a smoothly evolving spherical potential and an analytic
approximation for dynamical friction.  They neglected both interactions
between particles and the gravitational response of the rest of the galaxy
to those particles.  They selected accretion events from merger trees
for each of their 11 $M_{\mathrm{vir}} = 1.4 \times 10^{12}~M_{\odot}$
galaxies in a universe with $\Omega_m = 0.3$, $\Omega_{\Lambda}
= 0.7$, $\Omega_b h^2 = 0.024$, $h = 0.7$, and $\sigma_8 = 0.9$.
They populated the dark matter halos they simulated with star particles
using a simple star formation law and initialized their orbital planes
with random orientations.  To maximize the computational efficiency of
their simulations, they only simulated accretion events with significant
stellar content.  This approximation resulted in a satellite mass
floor of $M_{\mathrm{vir}} \sim 10^{8}~M_{\odot}$.  Given their lack
of a self-consistent potential, the \citet{bul05} models are less able
to accurately track the dynamical evolution of debris accreted before
its host galaxy's last major merger.  The inner halo is preferentially
formed from ancient accretion events, so the parametric potentials used
in the \citet{bul05} models may not accurately reproduce the phase-space
properties of the inner halo within 20 kpc.  The \citet{bul05} models
are further described in \citet{rob05} and \citet{fon06a,fon06b}.

The Via Lactea 2 dark matter halo \citep{die08,mad08,zem09,ras11} was
simulated using the parallel tree $N$-body code PKDGRAV2 in a universe
with $\Omega_m = 0.238$, $\Omega_{\Lambda} = 0.762$, $h = 0.73$,
and $\sigma_8 = 0.74$.  The simulation considered the formation of a
halo with $M_{\mathrm{vir}} = 1.9 \times 10^{12}~M_{\odot}$ at $z = 0$
resolved with $1.1 \times 10^9$ particles of mass $4,100~M_{\odot}$ and
gravitational softening length 40 pc.  The 1\% most tightly bound dark
matter particles in each subhalo with $M_{\mathrm{tid}} > 10^7~M_{\odot}$
at infall were tagged as star particles in post processing.  The stellar
mass and metallicity of each star particle was assigned based on
an empirical model that ensures that the remnant population of dSph
surviving to the present day reproduces the observed properties of the
Milky Way's satellite galaxies \citep{ras11}.  The Via Lactea 2 simulation
fully accounts for gravitational interactions between particles and
self-consistently determines the growth of the mass of the main halo, so
it better tracks ancient accretion events than the \citet{bul05} models.

\subsubsection{Description of Our Simulated Observations}

To simulate a SEGUE-like observation of one of the \citet{bul05} halos
or the Via Lactea 2 halo, we first place an observer at 8 kpc from the
center of the halo along the $x$-axis of the simulation.  This choice
is arbitrary, as there are no stellar disks in either simulation.
We consider only those star particles with $|z| > 4$ kpc that are more
than 10 kpc but less than 20 kpc from the center of the halo.  We compute
the radial velocity and Galactic coordinate of every star particle in
that volume as it would be viewed by the observer.  For every SEGUE line
of sight listed in Tables~\ref{tbl-1} and~\ref{tbl-2}, we compute the
range of $l$ and $b$ observed in the MPMSTO star population along that
line of sight.  We use that range of Galactic coordinates to select the
star particles from the simulations that would have been observed by SEGUE
along that line of sight.  As before, we compute the luminosity-weighted
mean metallicity of the stars in six distances ranges.  In that way,
we obtain line-of-sight-averaged metallicities that match the line of
sight averaged metallicities observed in the Milky Way's smooth halo
and presented in Tables~\ref{tbl-1} and \ref{tbl-2}.  One caveat is
that the metallicities reported in \citet{bul05} and \citet{ras11}
are calibrated to match the true iron abundances of dSph galaxies,
and not the total metallicities used in this analysis.

\begin{figure}
\epsscale{1.0} 
\plottwo{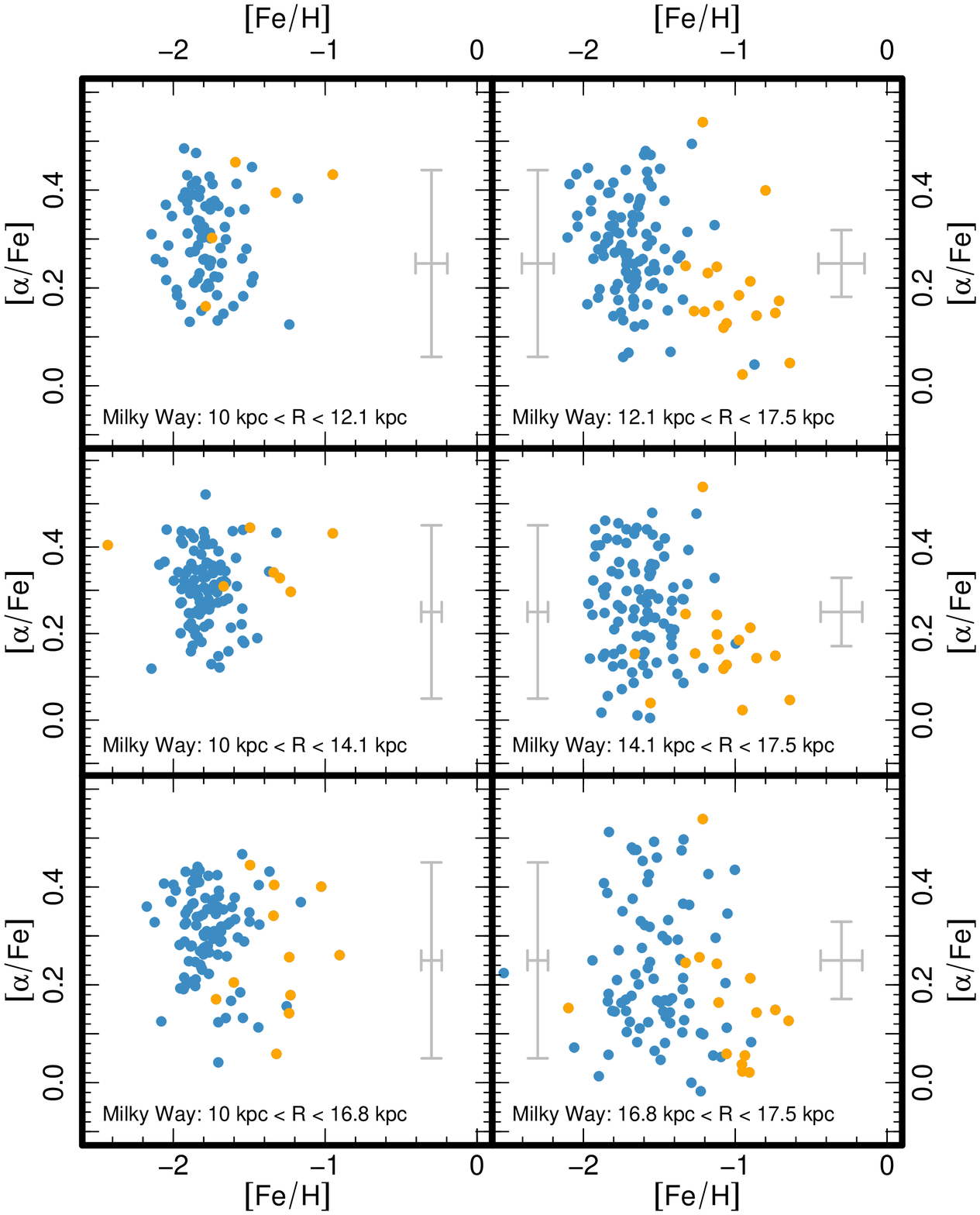}{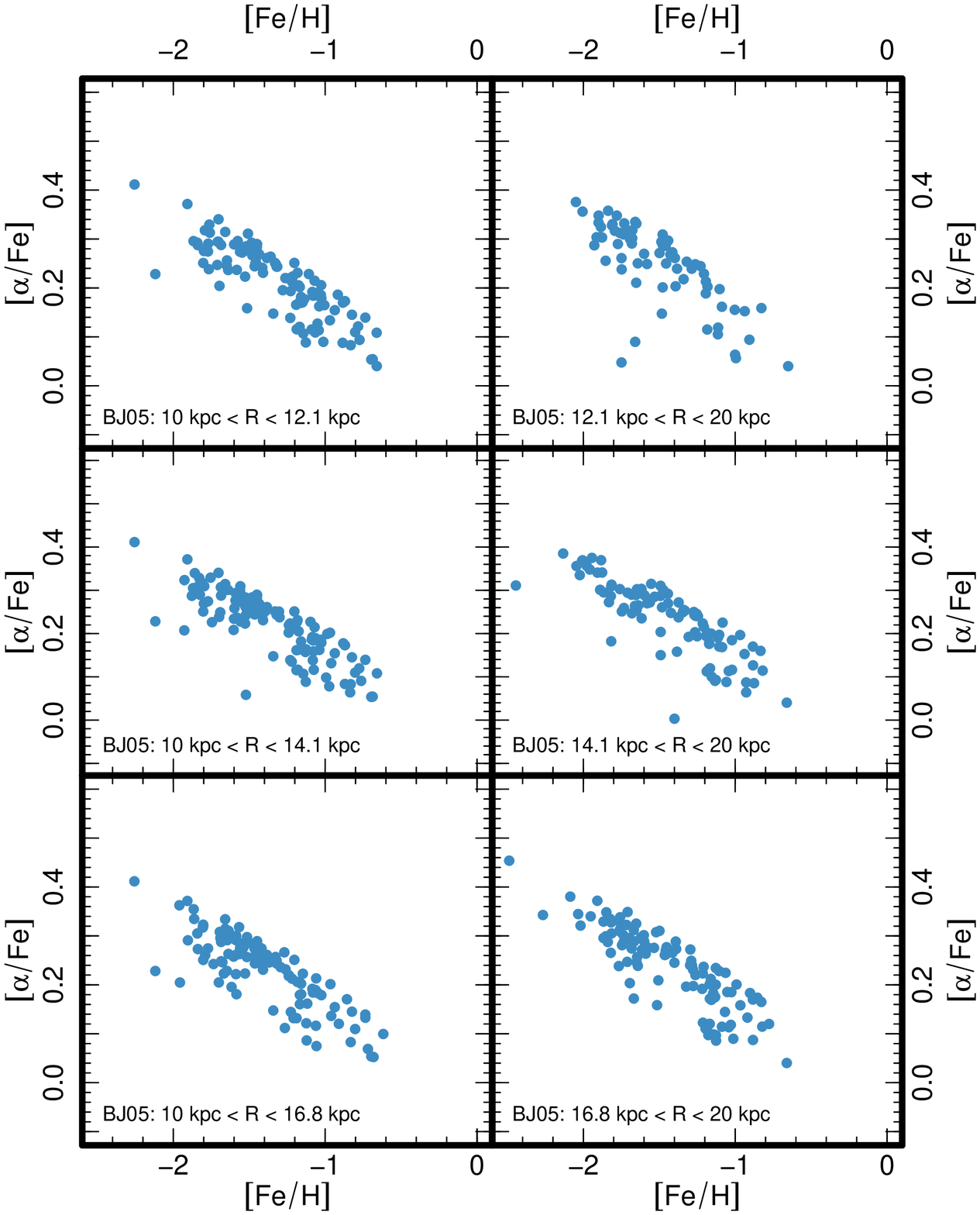}
\caption{Line of sight averaged abundance distribution as a function of
distance from the Galactic center in the smooth component of the halo in
the [Fe/H]--[$\alpha$/Fe] plane.  We split our tracer sample in half very
roughly in distance at three different split points: 12.1, 14.1 and 16.8
kpc.  Each row corresponds to the result for a single split-point.  In all
cases, the nearby half of the sample is in the left column and the distant
half of the sample is in the right column.  Left: from Tables~\ref{tbl-1}
and \ref{tbl-2}, the result for our pure smooth halo MPMSTO population.
The typical precision of our measurement is given by the gray error bars.
Our precision is a function of metallicity, so if there are two sets of
error bars present, the left bars apply for [Fe/H] $\lesssim -1.0$ while
the right bars apply for [Fe/H] $\gtrsim -1.0$.  We plot in blue lines
of sight with $|b| > 30^{\circ}$ and in orange lines of sight with $|b|
\leq 30^{\circ}$.  The smooth halo is dominated by the familiar $-1.8
\lesssim$ [Fe/H] $\lesssim -1.5$ and [$\alpha$/Fe]-enhanced population,
though at larger distances and low Galactic latitude the pure smooth
halo also has a dSph-like component with $\mathrm{[Fe/H]} \approx -1.0$
and [$\alpha$/Fe] $\approx 0.1$.  The feature at low Galactic latitude
is likely related to the low-latitude substructure first identified in
Monoceros \citep[e.g.,][]{new02,belo06b,jur08,ive08,schl11}.  Right: the
equivalent luminosity-weighted average abundance distribution that results
from a SEGUE-like observation of halo 10 from \citet{bul05}.  The result
is similar for all 11 halos presented in that paper.  We discuss the
significance of the observed differences between our observations of the
Milky Way and of the \citet{bul05} models in Section 3.2.\label{fig04}}
\end{figure}

In the right-hand panel of Figure~\ref{fig04}, we plot the result of
a SEGUE-like observation of halo 10 from \citet{bul05}.  Though the
\citet{bul05} models lack the gas-rich major mergers at high redshift
that \citet{rob05} advocated as the origin of the kinematically smooth
halo stellar population, they do illuminate the properties of a stellar
halo built entirely from disrupted satellite galaxies.  In an effort
to better isolate the smooth component of the \citet{bul05} models,
we attempted to find and remove ECHOS from those models.  However,
there are generally too few star particles to differentiate between
substructure and smooth component, and our search produced no results.
For that reason, we treat an entire halo from \citet{bul05} within 20
kpc of the center of the halo as a smooth component.  The star particle
tagging prescription applied to Via Lactea 2 did not track [$\alpha$/Fe],
so we cannot make the equivalent plot for Via Lactea 2.

Figure~\ref{fig04} indicates that the \citet{bul05} models are a poor
match to our observation of the smooth halo, though there is a hint
that they become a better match farther from the Galactic center.
In particular, the \citet{bul05} models have a significant [Fe/H]
$\approx -1.0$ and [$\alpha$/Fe] $\approx 0.1$ population in all distance
bins, while the [Fe/H] $\approx -1.0$ and [$\alpha$/Fe] $\approx 0.1$
population we observe in the Milky Way is confined to the region far
from the Galactic center and at low Galactic latitude.  Indeed, given
that our errors in average [Fe/H] are at most 0.2 dex, there is no way
to reconcile the left hand panels of both plots in Figure~\ref{fig04}
with measurement error alone.  If the low-latitude dSph-like composition
we observe in the smooth halo at large radius and low Galactic latitude
is somehow related to the stellar disk of the Milky Way, then it is
unreasonable to expect it to be reproduced by the \citet{bul05} models
(which have no stellar disk).

The absence of a significant [Fe/H] $\approx -1.0$ and [$\alpha$/Fe]
$\approx 0.1$ population in the smooth halo of the Milky Way relative to
the \citet{bul05} models has two main implications.  First, that most of
the Milky Way's halo was accreted at high redshift, and therefore at high
[$\alpha$/Fe] as inferred by \citet{rob05}.  Second, that the stellar mass
that has phased mixed into the smooth halo from massive substructures
(with high [Fe/H]) is dwarfed by the classical smooth halo component.
This observation is in contrast with the chemical abundances of ECHOS
studied in \citetalias{schl11}, where we found that more recently
accreted stars still kinematically distinct as ECHOS have both higher
[Fe/H] and lower [$\alpha$/Fe] than the smooth halo.

\begin{figure}
\plotone{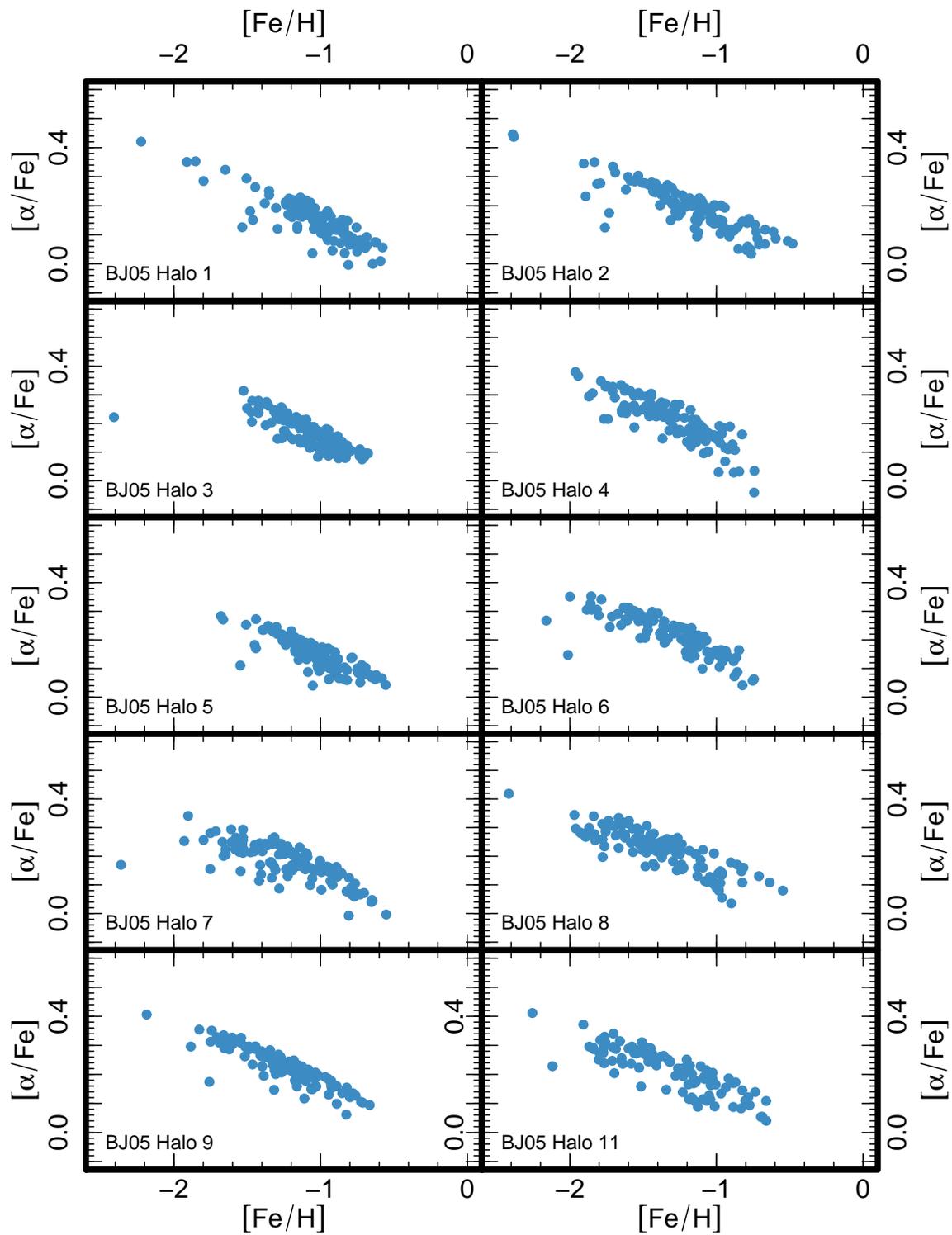}
\caption{Line of sight averaged abundance distribution in the
[Fe/H]--[$\alpha$/Fe] plane for the other 10 halos from \citet{bul05}.
We include only those star particles with $|z| > 4$ kpc that are more than
10 kpc but less than 20 kpc from the center of the halo.\label{fig05}}
\end{figure}

\begin{figure}
\plottwo{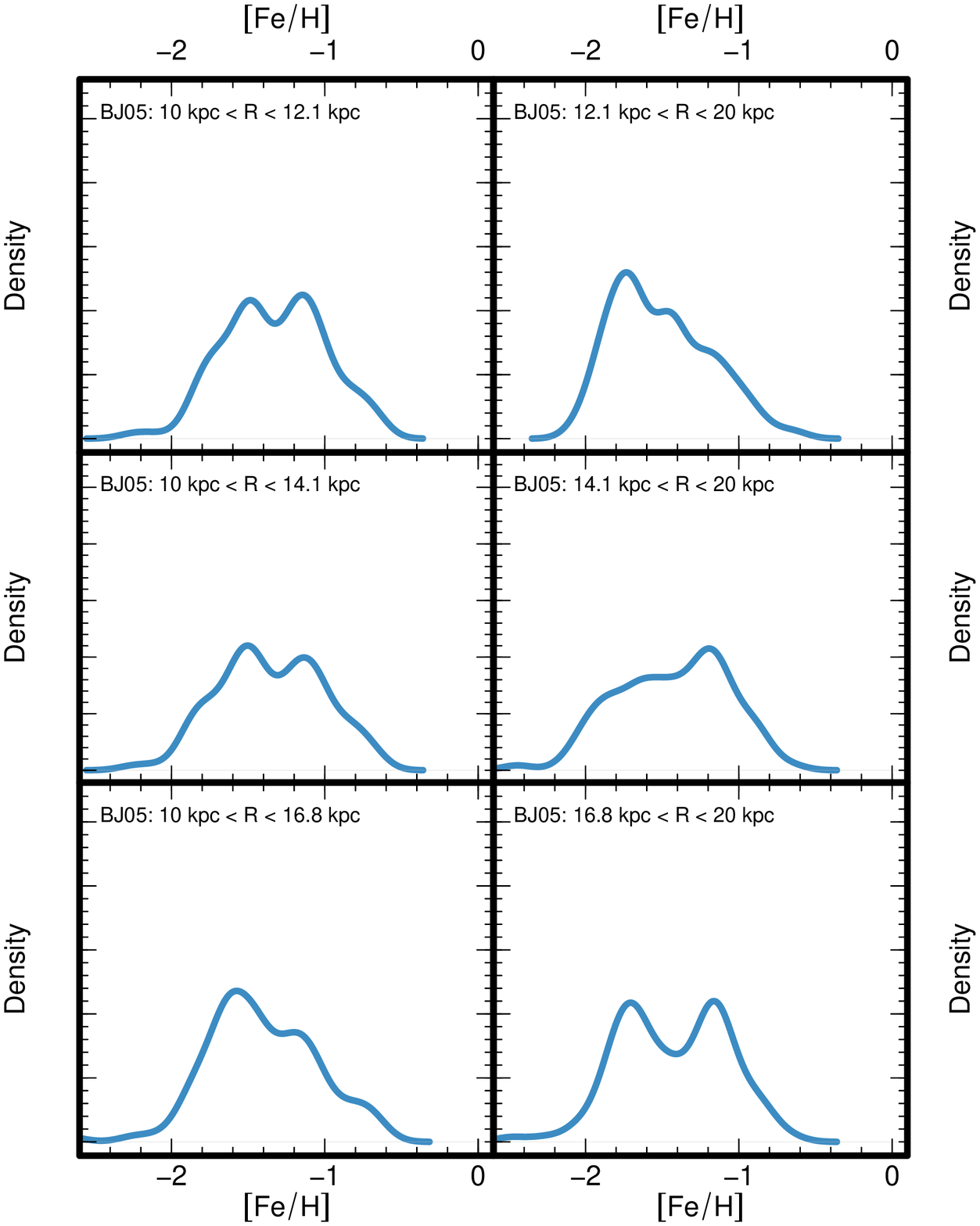}{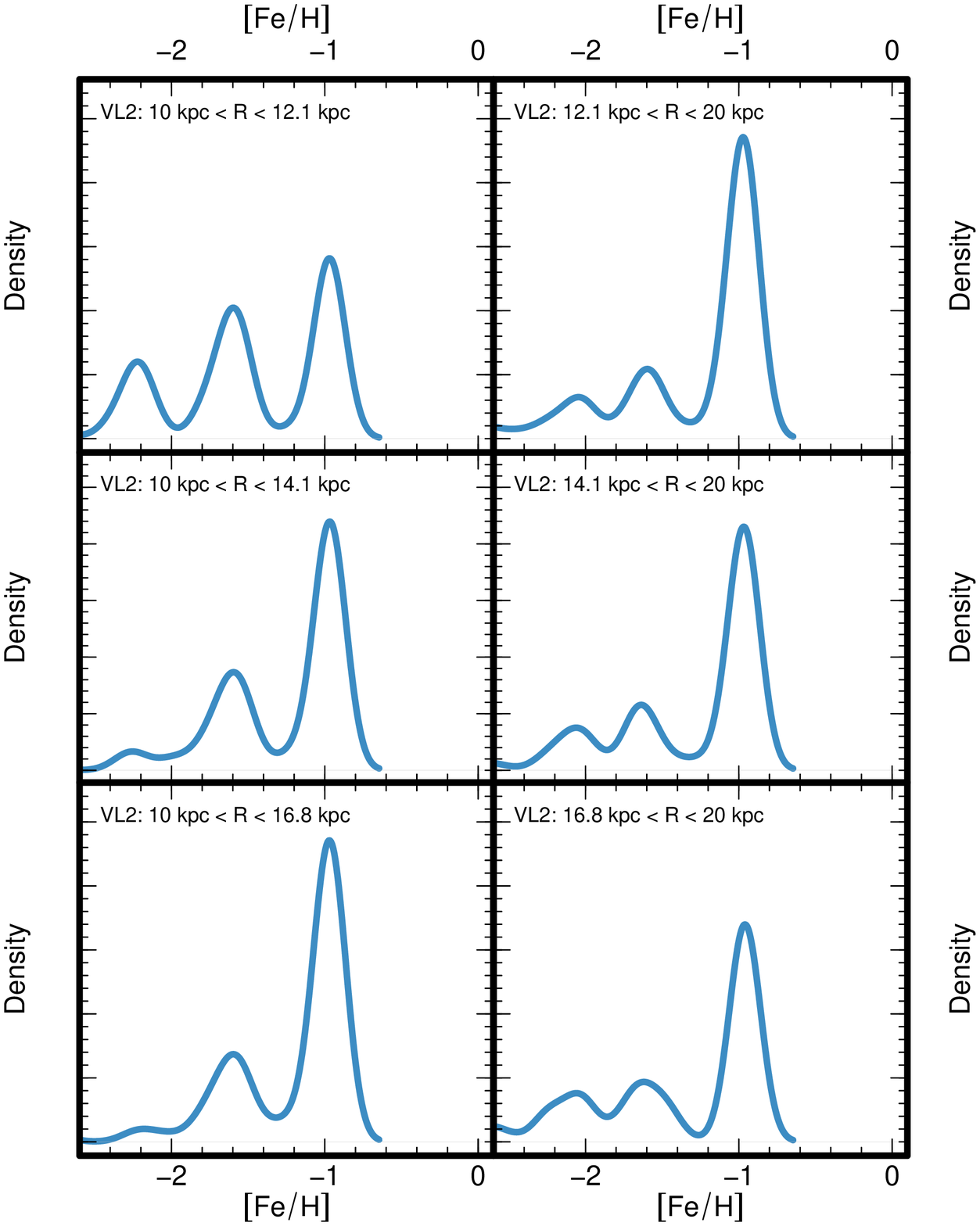}
\caption{Density of line-of-sight-averaged metallicity distribution
as a function of distance from the center of the halo kernel smoothed
to 0.1 dex.  Left: halo 10 from \citet{bul05}.  Right: the Via Lactea
2 halo \citep[e.g.,][]{die08,mad08,zem09}.  Dark matter particles in
the Via Lactea 2 halo were tagged as star particles with metallicities
determined by a simple satellite instantaneous enrichment at accretion
model described in \citet{ras11}.  The strongly peaked metallicity
distribution of the Via Lactea 2 halo is the result of the accretion
of two massive satellites with [Fe/H] $\approx -1.0$ at high redshift.
Neither dark matter only $N$-body simulation is a good match for our
observations of the metallicity of the smooth halo, indicating the
importance of gas and in situ star formation in the origin of the smooth
halo within 17.5 kpc of the Galactic center.\label{fig06}}
\end{figure}

Figure~\ref{fig05} shows the [Fe/H]--[$\alpha$/Fe] morphology of the
other 10 halos from \citet{bul05}.  While each of the 11 halos from
\citet{bul05} has a unique accretion history, the [Fe/H]--[$\alpha$/Fe]
morphology present in Figure~\ref{fig04} is common to all 11 halos.
We plot the metallicity distributions of both halo 10 from \citet{bul05}
and the Via Lactea 2 halo in Figure~\ref{fig06}.  The strongly-peaked
metallicity distribution we observe in the inner 20 kpc of the Via
Lactea 2 halo is the result of two massive satellites accreted at high
redshift.  These two satellites contributed 80\% of the stellar mass
in that volume and had [Fe/H] $= -0.94$ and [Fe/H] $= -1.04$.  While we
showed that we are capable of accurately measuring [Fe/H] near [Fe/H] $=
-1$ in \citetalias{schl11}, our observational data are inconsistent
with a halo having a large fraction of its stars at [Fe/H] $= -1$ as
suggested by Figure~\ref{fig06}, even after modeling the metallicity
bias of the UV excess selection.

\subsection{Spatial Autocorrelation}

Tables~\ref{tbl-1} and~\ref{tbl-2} give the spatial distribution of [Fe/H]
and [$\alpha$/Fe] in the kinematically smooth halo MPMSTO population.
As we showed in Figure~\ref{fig04}, the low-latitude substructure at
[Fe/H] $\approx -1.0$ and [$\alpha$/Fe] $\approx 0.1$ is very prominent
at low Galactic latitude.  At the same time, Figure~\ref{fig04} does
not reveal whether lines of sight that are close together on the sky
also have similar metallicities.  As we described in Section 1, a halo
that was formed in a dissipative process will not have the property that
lines of sight that are close together on the sky also have similar
metallicities.  On the other hand, a halo that was formed from disrupted
satellite galaxies will have the property that lines of sight that are
close together on the sky also have similar metallicities.

To determine whether lines of sight that are close together on the
sky also have similar metallicities, we use Moran's $I$ statistic
\citep{mor50} to quantify the spatial autocorrelation of the chemical
abundances:

\begin{eqnarray}\label{eq01}
I & = & \frac{n}{\sum_{i}^{n} \sum_{j}^{n} w_{i,j}} \frac{\sum_{i}^{n} \sum_{j}^{n} w_{i,j} (x_{i} - \overline{X}) (x_{j} - \overline{X})}{\sum_{i}^{n} (x_{i} - \overline{X})^2}.
\end{eqnarray}

\noindent
In Equation~\ref{eq01}, $n$ is the number of lines of sight, $x_{i}$ is
average metallicity inferred for a given line of sight, $\overline{X}$ is
the mean metallicity of the sample, and $w_{i,j}$ is the spatial weight.
In this case, each $w_{i,j}$ is the inverse of the angular distance
between line of sight $i$ and line of sight $j$.  The expected value
and variance of Moran's $I$ under the null hypothesis of no spatial
variation are \citep[e.g.,][]{cli81,wal04}:

\begin{eqnarray}\label{eq02}
\mathrm{E}(I) & = & -\frac{1}{n-1}, \\
\mathrm{Var}(I) & = & \frac{n^2 S_{1} - n S_{2} + 3 S_{0}^2}{(n-1) (n+1) S_{0}^2} - \left(\frac{1}{n-1}\right)^2,
\end{eqnarray}

\noindent
where

\begin{eqnarray}\label{eq03}
S_{0} & \equiv & \sum_{i}^{n} \sum_{j}^{n} w_{i,j}, \\
S_{1} & \equiv & \frac{1}{2 \sum_{i}^{n} \sum_{j}^{n} (w_{i,j} + w_{j,i})^2}, \\
S_{2} & \equiv & \sum_{i}^{n} \left(w_{i+} + w_{+j}\right)^2, \\
w_{i+} & \equiv & \sum_{j}^{n} w_{i,j}, \\
w_{+j} & \equiv & \sum_{j}^{n} w_{j,i}.
\end{eqnarray}

Positive values of Moran's $I$ statistic indicate positive spatial
autocorrelation, while negative values indicate negative spatial
autocorrelation.  In other words, if lines of sight that are close
together on the sky are significantly more likely to have similar
metallicities than expected based on chance, then Moran's $I$ statistic
will be significantly positive.

Though Moran's $I$ statistic gives a good indication of the existence
of spatial autocorrelation in data, it does not indicate the angular
scale at which spatial autocorrelation is strongest.  To constrain the
spatial scale of autocorrelation, we define the statistic $\Theta_{A,i}$:

\begin{eqnarray}\label{eq04}
\Theta_{A,i} & = & \frac{1}{m} \sum_{j \in A, j \not= i} \left(x_{i} - x_{j} \right)^2,
\end{eqnarray}

\noindent
where $x_{i}$ and $x_{j}$ are the average metallicity inferred for the
lines of sight indexed by $i$ and $j$, $A$ is the set of lines of sight
within a given angular scale of $x_{i}$, and $m$ is the number of lines
of sight in $A$.

For each line of sight in Tables~\ref{tbl-1} and~\ref{tbl-2},
we compute $\Theta_{A,i}$ at each of ten angular scales
$10^{\circ},20^{\circ},\ldots,100^{\circ}$.  To estimate both $\Theta_{A}$
and our uncertainty in $\Theta_{A}$ at each angular scale, we use a
Monte Carlo simulation.  Instead of using the set $A$ of all lines
of sight within a given angular scale of the line of sight denoted by
$x_{i}$, we create a bootstrap set $A'$ by sampling with replacement
$m$ lines of sight from all $m$ available in $A$.  We compute
$\Theta_{A',i}$ in this way 100 times for each line of sight and each
angular scale.  We then compute the mean $\mathrm{E}(\Theta_{A',i})$
and variance $\mathrm{Var}(\Theta_{A',i})$ of $\Theta_{A',i}$ for
each line of sight and each angular scale over the 100 bootstrap
realizations of $A'$.  We then compute $\mathrm{E}(\Theta_{A})$ by
averaging $\mathrm{E}(\Theta_{A',i})$ over all $n$ lines of sight and
$\mathrm{Var}(\Theta_{A})$ by averaging $\mathrm{Var}(\Theta_{A',i})$
over all $n$ lines of sight and dividing by $\sqrt{n}$.

To determine the distribution of $\Theta_{A}$ under the null hypothesis
that there is no spatial autocorrelation in the data, we use a second
Monte Carlo simulation.  We compute $\Theta$ as before, though instead
of using the set $A$ of all lines of sight within a given angular scale
of the line of sight denoted by $x_{i}$, we randomly sample $m$ lines of
sight from all $n$ available regardless of spatial proximity and call
this set $B'$.  We compute $\Theta_{B',i}$ in this way 100 times for
each line of sight and each angular scale.  We then compute the mean
$\mathrm{E}(\Theta_{B',i})$ and variance $\mathrm{Var}(\Theta_{B',i})$
of $\Theta_{B,i}$ for each line of sight and each angular scale over
the 100 bootstrap realizations of $B'$.  We then compute the mean
$\mathrm{E}(\Theta_{B})$ and variance $\mathrm{Var}(\Theta_{B})$
under the null hypothesis of no spatial autocorrelation by averaging
$\mathrm{E}(\Theta_{B',i})$ over all $n$ lines of sight and by averaging
$\mathrm{Var}(\Theta_{B',i})$ over all $n$ lines of sight and dividing
by $\sqrt{n}$.

The significance of spatial autocorrelation at a given angular scale
is then:

\begin{eqnarray}\label{eq05}
\Omega & = & \frac{\mathrm{E}(\Theta_{B}) - \mathrm{E}(\Theta_{A})}{\sqrt{\mathrm{Var}(\Theta_{A}) + \mathrm{Var}(\Theta_{B})}}.
\end{eqnarray}

\begin{figure}
\plottwo{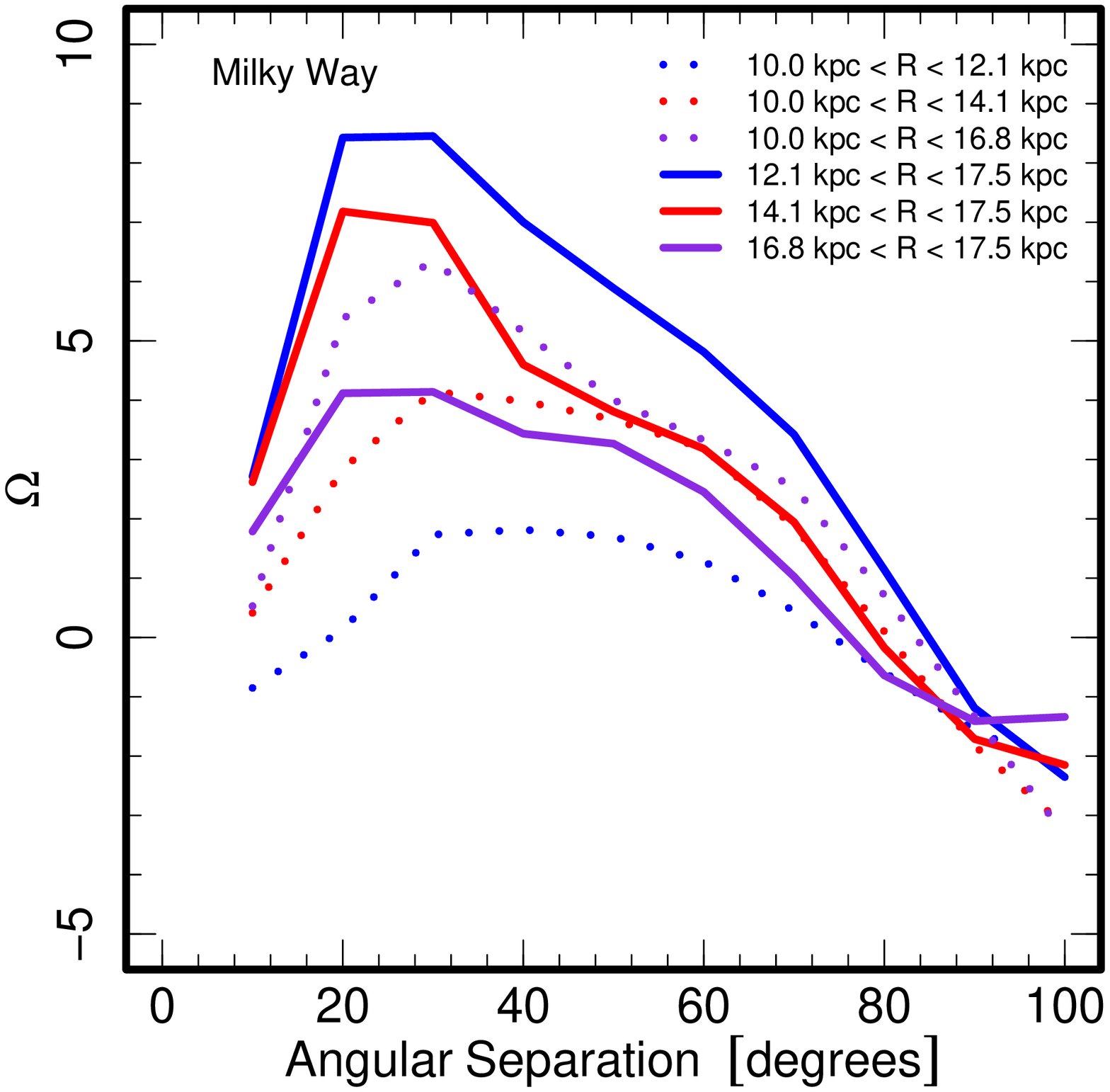}{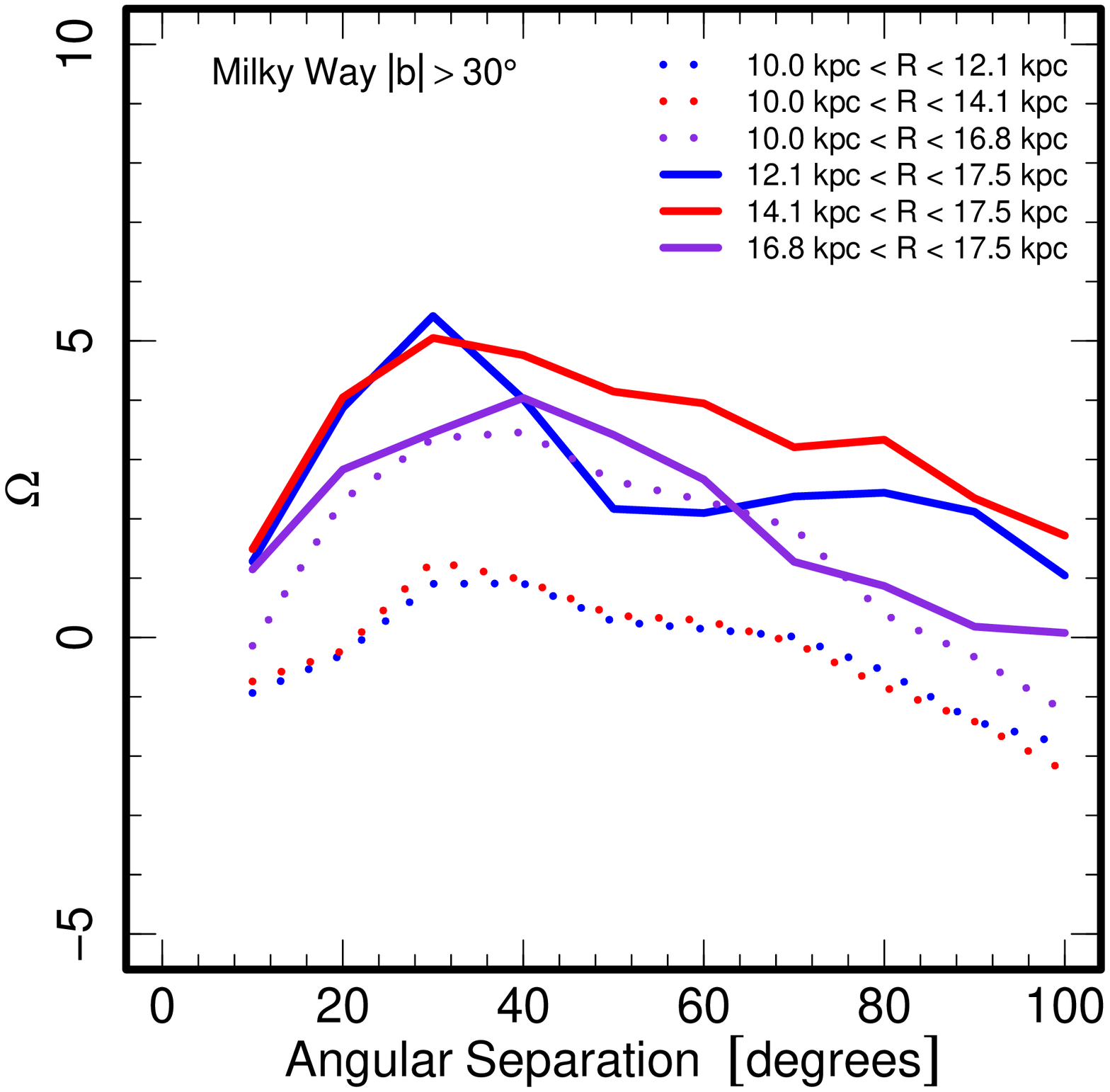}
\caption{Quantification of spatial autocorrelation in [Fe/H] in
the Galaxy's pure smooth halo MPMSTO population as a function of
Galactocentric distance and angular scale.  Positive values of the
statistic $\Omega$ indicate spatial autocorrelation.  Left: the result
for all lines of sight.  Right: the result for lines of sight with $|b|
> 30^{\circ}$.  Ignoring the low-latitude substructure and looking
only at high Galactic latitude, there is no spatial autocorrelation
in [Fe/H] interior to 14.1 kpc.  Beyond 14.1 kpc however, there is
significant spatial autocorrelation in [Fe/H] with characteristic
angular scale between 20$^{\circ}$ and 30$^{\circ}$.  At 14 kpc -- the
median Galactocentric distance of the MPMSTO stars in our sample -- that
corresponds to a physical scale of approximately 5 kpc.  This observation
suggests that the relative contribution of disrupted satellite galaxies
to the stellar population of the smooth halo increases with distance,
becoming comparable to the classical smooth halo component at about
15 kpc.\label{fig07}}
\end{figure}

\begin{figure}
\plottwo{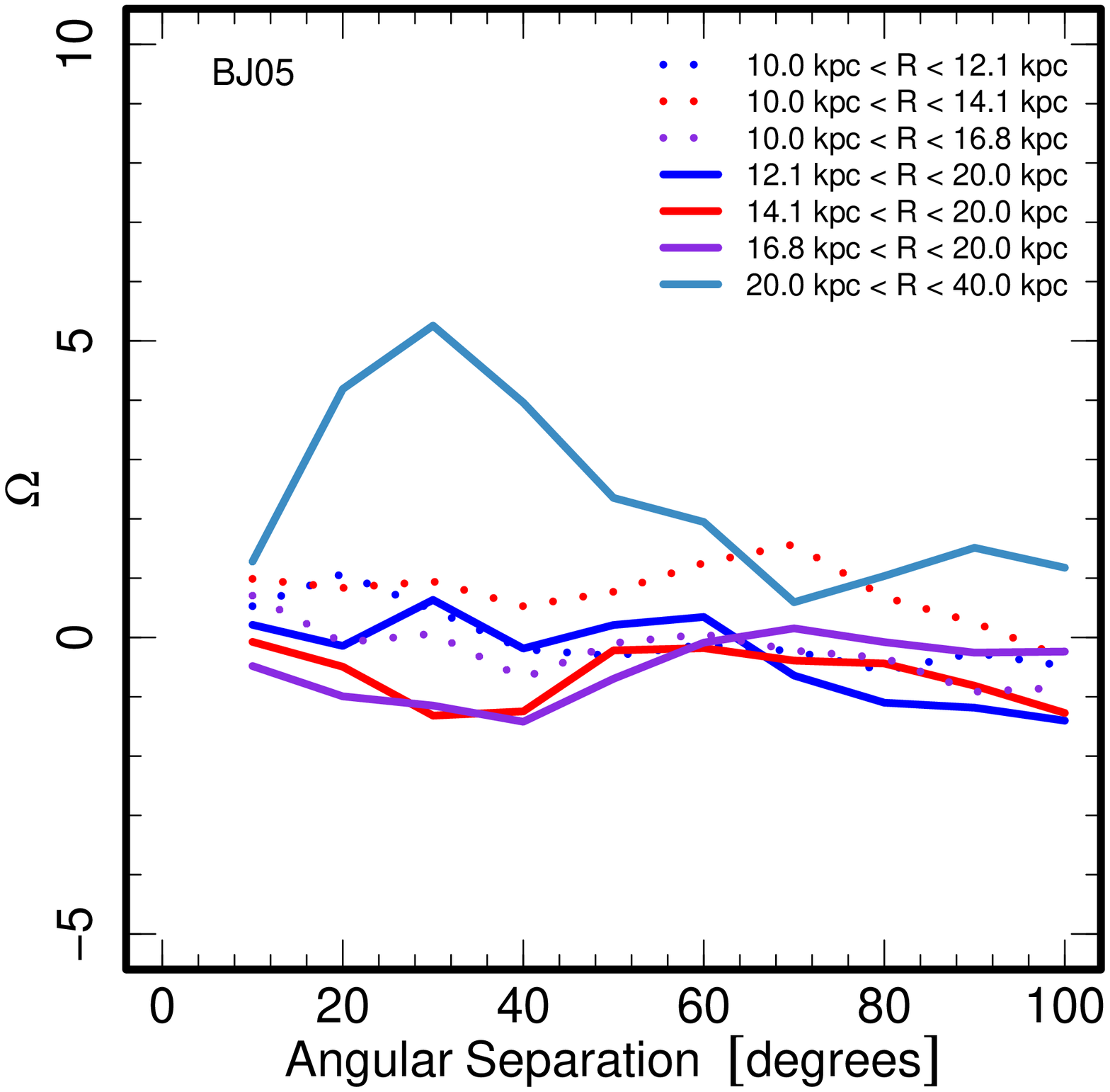}{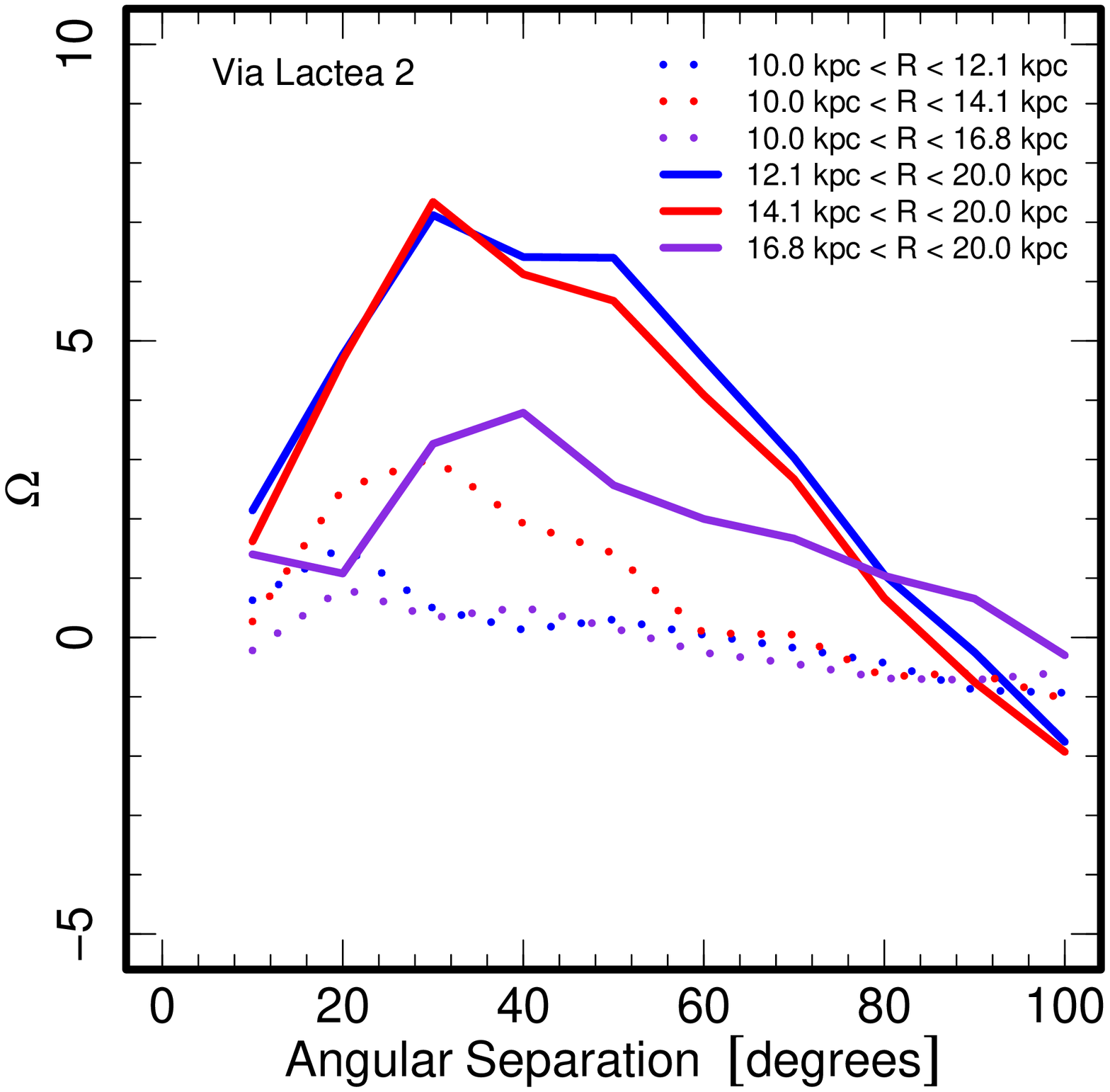}
\caption{Quantification of spatial autocorrelation in [Fe/H]
in simulations of Milky Way analog formation.  Left: halo 10 from
\citet{bul05}.  Right: the Via Lactea 2 halo.  There is no significant
spatial autocorrelation in [Fe/H] within 20 kpc of the center of halo
10 from \citet{bul05}.  There does appear to be significant spatial
autocorrelation in [Fe/H] further from the center of the halo, where the
\citet{bul05} simulations likely model halo formation more accurately.
On the other hand, there is significant spatial autocorrelation in [Fe/H]
within 20 kpc of the center of the Via Lactea 2 halo.  The appearance
of significant spatial autocorrelation in [Fe/H] with 20 kpc of the halo
center and the angular scale of that spatial autocorrelation in the Via
Lactea 2 halo are very similar to the same quantities measured in the
halo of the Milky Way.  The rapid drop-off in the significance of spatial
autocorrelation in [Fe/H] at large angular scale in the Via Lactea 2
halo may indicate that there is less large-scale diffuse substructure
in the Via Lactea 2 halo than in the Milky Way.  The extra large-scale
substructure in the Milky Way may be related to the ongoing merger with
the Sagittarius dwarf galaxy.\label{fig08}}
\end{figure}

\noindent
Like Moran's $I$ statistic, positive values of $\Omega$ indicate spatial
autocorrelation; negative values of $\Omega$ indicate spatial dispersion.
We report Moran's $I$ statistic for the data in Table~\ref{tbl-1}
in Tables~\ref{tbl-4} and~\ref{tbl-5}; we report the importance of
spatial autocorrelation graphically in Figure~\ref{fig07}.  We report
Moran's $I$ statistic for the \citet{bul05} halos and Via Lactea 2 in
Tables~\ref{tbl-6} and~\ref{tbl-7}, respectively; we report the importance
of spatial autocorrelation graphically in Figure~\ref{fig08}.  We find no
significant spatial autocorrelation in [$\alpha$/Fe], likely because our
measurement of that quantity is much less precise than our measurement of
[Fe/H].  For that reason, we do not consider [$\alpha$/Fe] any further.

We find that there is significant spatial autocorrelation in [Fe/H] in
the Milky Way beyond about 12 kpc from the Galactic center, indicated
by the very small $p$-values in Table~\ref{tbl-4} and large positive
values of $\Omega$ in the left-hand panel of Figure~\ref{fig07}.  The
spatial autocorrelation in [Fe/H] close to the Galactic center is very
likely due to the prominent substructure at low Galactic latitude with
[Fe/H] $\approx -1.0$ and [$\alpha$/Fe] $\approx 0.1$.  Indeed, if we
restrict our sample only to lines of sight with Galactic latitude $|b|
> 30^{\circ}$, the signal of spatial autocorrelation in [Fe/H] closer
to the Galactic center disappears.  This absence is indicated by the
large $p$-values in Table~\ref{tbl-5} and the small values of $\Omega$
in the right-hand panel of Figure~\ref{fig07}.

Even at $|b| > 30^{\circ}$ (i.e., ignoring the low latitude substructure),
there is still significant spatial autocorrelation in [Fe/H] far from
the center of the Galaxy in the classical smooth halo $-1.8 \lesssim$
[Fe/H] $\lesssim -1.5$ and [$\alpha$/Fe]-enhanced population.  Given the
poor precision of our MPMSTO star distance estimator, our constraint
on the distance at which spatial autocorrelation becomes apparent
is weak.  Nevertheless, the $p$-values from Moran's $I$ statistic in
Table~\ref{tbl-5} indicate that the effect becomes apparent beyond
about 14 kpc but before about 16 kpc.  The effect is very strong at
the largest distances probed by or MPMSTO star sample.  The smallest
angular scale that we are sensitive to spatial autocorrelation in [Fe/H]
is 10$^{\circ}$.  Interestingly, in Figure~\ref{fig07} the statistic
$\Omega$ peaks between 20$^{\circ}$ and 30$^{\circ}$.  This is the
characteristic angular scale of spatial autocorrelation.  At 14 kpc --
the median Galactocentric distance of the MPMSTO stars in our sample --
that corresponds to a physical scale of approximately 5 kpc.  This is much
larger than the scale of the disrupted satellite galaxies that contributed
the stellar populations that are likely the source of this effect.

After we identified the characteristic angular scale of spatial
autocorrelation in metallicity, we went back to the data in
Table~\ref{tbl-1} and computed Moran's $I$ statistic in 30$^{\circ}$
windows around each line of sight.  While we found that a handful of lines
of sight displayed significant spatial autocorrelation in their own local
30$^{\circ}$ window, the $p$-values that resulted from those observations
were orders of magnitudes larger than the $p$-values we computed for
the full sample of 118 lines of sight.  Many lines of sight display no
significant spatial autocorrelation in their own 30$^{\circ}$ window.
Consequently, the signal of spatial autocorrelation that we observe is not
the result of a small number of discrete locations in the sky.  Instead,
the significant spatial autocorrelation in [Fe/H] we observe is coming
from everywhere in the Galactic halo.

We also recomputed Moran's $I$ statistic using the signal to
noise-weighted line-of-sight average metallicities created by averaging
DR8 SSPP metallicities for all individual stars in the same lines of
sight and volumes described in Table~\ref{tbl-1}.  That analysis produced
a qualitatively similar result to that described above, albeit at much
lower statistical significance.

We do not identify any significant spatial autocorrelation in [Fe/H] in
the Milky Way analog halos from \citet{bul05} in the volume between 10
and 20 kpc from the halo center.  This lack of spatial autocorrelation
is indicated by the large $p$-values in Table~\ref{tbl-6} and small
values of $\Omega$ in the left-hand panel of Figure~\ref{fig08}.
The large $p$-values and small $\Omega$ values we present for halo
10 from \citet{bul05} are representative of the values we obtain from
equivalent observations of the other ten halos presented in \citet{bul05}.
The smoothness of the star-particle distribution near the center of the
halo could be physical, or it could be a byproduct of the limitations
of the simulations.  We do observe significant spatial autocorrelation
in the \citet{bul05} halos beyond 20 Galactocentric kpc though, where
the \citet{bul05} simulations likely model halo formation more accurately.

On the other hand, we do observe significant spatial autocorrelation in
[Fe/H] in the Via Lactea 2 halo in the volume between 10 and 20 kpc from
the halo center.  The significant spatial autocorrelation in [Fe/H] is
indicated by the small $p$-values in Table~\ref{tbl-7} and large values
of $\Omega$ in the right-hand panel of Figure~\ref{fig08}.  Since we
observe significant [Fe/H] spatial autocorrelation in the inner regions
of the Via Lactea 2 halo, but little [Fe/H] spatial autocorrelation in
the inner regions of any of the 11 halos from \citet{bul05}, the Via
Lactea 2 observation is not likely due to its unique accretion history.

The \citet{bul05} halos and the Via Lactea 2 halo have similar mass
and were simulated using similar cosmological parameters.  One small
difference between the two simulations is the value of $\sigma_8$, as
\citet{bul05} used $\sigma_8 = 0.9$ while Via Lactea 2 used $\sigma_8
= 0.74$.  While this affects the abundance of Milky Way analog halos,
it should not much affect the properties of individual Milky Way analog
halos \citep[e.g.,][]{fon11}.  As a result, the two simulations should not
have statistically disparate accretion histories.  For that reason, the
fact that we do not observe significant spatial autocorrelation in [Fe/H]
in the inner regions of any of the 11 halos presented in \citet{bul05},
but do observe the effect in the inner region of the Via Lactea 2 halo,
indicates that the discrepancy is not likely to result from differences
in halo merger history.

The observed differences between the two calculations may be a result
of the higher resolution and cosmologically self-consistent potential
of the more modern Via Lactea 2 simulation.  Another hypothesis is that
accretion along a filament in the Via Lactea 2 halo produced the observed
spatial autocorrelation, as the orbital planes of infalling satellites
in the \citet{bul05} models were chosen randomly and independently.
The simulations necessary to verify the latter speculative hypothesis
would be very computational intensive, and therefore are beyond the
scope of this analysis.

\section{Discussion}

We have three observations that a complete theoretical model for the
formation of the inner halo must reproduce: (1) the halo metallicity
distribution, (2) the joint halo [Fe/H]-[$\alpha$/Fe] distribution, and
(3) the significance of spatial autocorrelation in [Fe/H] in the halo.
The \citet{bul05} models to not reproduce any of the three measurements,
as expected by those authors because of their simplifying assumptions.
The Via Lactea 2 model does match (3) even as it fails to reproduce (1)
and (2).  While self-consistent, high-resolution pure-accretion models may
be able reproduce (3), they have not yet been shown capable of reproducing
(1) and (2).

As described in Section 3.3, we do not observe significant spatial
coherence in [Fe/H] in the halo of Milky Way inside of 15 kpc.
This observation is consistent with both a smooth halo formed
through a combination of in situ star formation and dissipative major
mergers at high redshift and a smooth halo formed through the phase
mixing of the debris of disrupted satellite galaxies.  However, the
\citet{bul05} halos also simulate the chemical abundance structure
of a halo built entirely from accreted galaxies.  As we observed in
Figures~\ref{fig04}-\ref{fig06}, the overall metallicity of such
halos are incompatible with the overall metallicity of the halo
of the Milky Way.  As a result, our observation of the chemical
structure and spatial autocorrelation in [Fe/H] in the smooth
halo supports the idea of in situ star formation and dissipative
major mergers at high redshift as the origin of the smooth halo
\citep[e.g.,][]{rob05,zol09,zol10,fon11,tis11,mcc11}.

We observe significant spatial autocorrelation in [Fe/H] in the halo
of Milky Way exterior to about 15 Galactocentric kpc.  The spatial
autocorrelation is a global property of the entire halo beyond 15 kpc,
and it cannot be simply associated with a few locations on the sky.
These observations favor a halo model in which the relative contribution
of accreted galaxies to the stellar population of the halo increases
with radius, becoming observable in our data relative to the smooth halo
population at about 15 kpc.

The dominance of the classical smooth halo close to the center of
the Galaxy does not mean that disrupted satellite galaxies have not
contributed to the halo stellar population at small distance.  The most
massive galaxies accreted by the Milky Way will quickly spiral into
the center of the halo due to dynamical friction and deposit most of
their stars in the inner halo.  The chemical abundances of the recently
accreted stars in the inner halo indicate that they were likely formed
in the most massive of the satellite galaxies recently accreted by the
Milky Way \citep[e.g.,][]{fon06b,schl11}.  Consequently, there is likely
a substantial number of accreted stars close to the Galactic center.
However, the classical smooth component of the halo peaks close to
the Galactic center too, dominating the component of the halo from
tidally-disrupted satellite galaxies.

Though the densities of both the classical smooth halo and its
tidally-disrupted satellite galaxy counterpart decrease with distance, our
observation that there is significant spatial autocorrelation in [Fe/H]
far from the center of the Galaxy indicates that the classical smooth
component decreases more quickly with distance than the tidally-disrupted
satellite galaxy component.  Our conclusion is that 15 kpc from the center
of the Galaxy is the approximate radius at which the accreted halo becomes
non-negligible in comparison to the halo component formed in a combination
of in situ star formation and dissipative major mergers at high redshift.
While our observation of spatial autocorrelation in [Fe/H] indicates that
the contribution of accretion to the smooth component becomes important
at about 15 kpc from the Galactic center, it is difficult to quantify.
In \citetalias{schl09}, we found that about 30\% of the inner halo volume
has 10\% of its MPMSTO population in diffuse ECHOS.  We speculate that
this diffuse ECHOS population may be the origin of the signal we observe.

Our observation of spatial autocorrelation in [Fe/H] is a first example
of in situ statistical chemical tagging in the halo.  Our result suggests
that future surveys of the halo stellar population using high-resolution
multi-object spectroscopy will identify dynamically ancient groups of
stars with similar abundance patterns indicative of formation in a single
star forming region.  In particular, future surveys with the ability to
chemically tag stars in situ more than 15 kpc from the Galactic center
should find many prominent chemical substructures with angular scales
comparable to $30^{\circ}$.

The observation that the classical smooth halo component ceases to be the
dominant component of the stellar population of the halo at about 15 kpc
from the Galactic center is consistent with the dual-halo hypothesis
advanced in \citet{caro07,caro10} and \citet{bee12}.  Though we do
not observe a significant metallicity gradient in the smooth halo, our
observations do not preclude the existence of a metallicity gradient.
Likewise, this result is analogous to the \citet{dej10} observation that
the halo density profile falls off rapidly with distance from the Galactic
center to a distance of about 15 to 20 kpc.  Beyond that distance, they
found a substantially lower density, slowly varying halo density profile.
\citet{ses11} observed something similar at even larger radius.

\section{Conclusion}

We find significant spatial coherence in [Fe/H] in the MPMSTO population
in the stellar halo of the Milky Way beyond about 15 kpc from the
center of the Galaxy.  That spatial coherence suggests that the relative
contribution of disrupted satellite galaxies to the stellar population
of the smooth halo increases with radius, becoming observable in our data
relative to the classical kinematically smooth halo at about 15 kpc from
the Galactic center.  SEGUE-like observations of the Via Lactea 2 halo
indicate that spatial autocorrelation in [Fe/H] is a generic feature of
stellar halos formed entirely by accretion.  We find that this spatial
autocorrelation is strongest on angular scales between 20$^{\circ}$
and 30$^{\circ}$, corresponding to a physical scale of about 5 kpc at
15 kpc.  Though we find that a significant fraction of the stellar halo
of the Milky Way beyond about 15 Galactocentric kpc is likely comprised
of the phase-mixed debris of satellite galaxies, the morphology of the
halo in the [Fe/H]--[$\alpha$/Fe] plane inside of 15 kpc is not well
matched by phase-mixed tidal debris.  Rather, the smooth halo inside of
15 kpc is likely formed through a combination of in situ star formation
and dissipative major mergers at high redshift.

\acknowledgments We thank Jim Gunn, Raja Guha Thakurta, and Don Schneider
for useful comments.  We are grateful to the anonymous referee for many
insightful comments that improved this paper.  This research has made use
of NASA's Astrophysics Data System Bibliographic Services.  This material
is based upon work supported under a National Science Foundation (NSF)
Graduate Research Fellowship.  CMR gratefully acknowledges support
from the David and Lucile Packard Foundation.  YSL and TCB acknowledge
partial support for this work from PHY 02-16783 and PHY 08-22648:
Physics Frontiers Center / Joint Institute for Nuclear Astrophysics
(JINA), awarded by the NSF.  PM acknowledges support for this work from
the NSF through grants AST-0908910 and OIA-1124453.  Funding for the
SDSS and SDSS-II has been provided by the Alfred P. Sloan Foundation,
the Participating Institutions, the NSF, the U.S. Department of
Energy, the National Aeronautics and Space Administration, the Japanese
Monbukagakusho, the Max Planck Society, and the Higher Education Funding
Council for England. The SDSS Web Site is http://www.sdss.org/.

The SDSS is managed by the Astrophysical Research Consortium for the
Participating Institutions. The Participating Institutions are the
American Museum of Natural History, Astrophysical Institute Potsdam,
University of Basel, University of Cambridge, Case Western Reserve
University, University of Chicago, Drexel University, Fermilab, the
Institute for Advanced Study, the Japan Participation Group, Johns
Hopkins University, the Joint Institute for Nuclear Astrophysics, the
Kavli Institute for Particle Astrophysics and Cosmology, the Korean
Scientist Group, the Chinese Academy of Sciences (LAMOST), Los Alamos
National Laboratory, the Max-Planck-Institute for Astronomy (MPIA),
the Max-Planck-Institute for Astrophysics (MPA), New Mexico State
University, Ohio State University, University of Pittsburgh, University
of Portsmouth, Princeton University, the United States Naval Observatory,
and the University of Washington.

{\it Facility:} \facility{Sloan}

\appendix

\section{A. Description of Co-addition Algorithm} 

Using the co-addition technique described in detail in
\citetalias{schl11}, for each subsample we create the ``average" MPMSTO
spectrum for that subsample by creating a single co-added spectrum from
all of the MPMSTO spectra in that subsample.  We include in the co-added
spectra only those spectra that correspond to MPMSTO stars within the 500
K range in effective temperature $T_{\mathrm{eff}}$ that maximizes the S/N
noise in the resultant co-added spectrum.  We shift each MPMSTO spectrum
eligible for inclusion in a co-add to a heliocentric radial velocity
$v_r = 0$ km s$^{-1}$.  We then use natural cubic spline interpolation to
interpolate both the spectrum and its inverse variance on to a common grid
in wavelength.  Next, we numerically integrate the area under the curve
defined by the spectrum and normalize both the spectrum (by dividing by
the normalization factor) and the inverse variance (by multiplying by
the normalization factor squared) to ensure that each spectrum that is
to be included in the co-add has the same scale.  For each population
of interest, we then create an ensemble of realizations of the co-added
spectrum by bootstrap resampling from the set of radial-velocity zeroed,
interpolated, and normalized spectra that belong to that population.
Each spectrum contributes to each wavelength bin in proportion to its
inverse variance in that bin relative to the other spectra selected for
co-addition.  We tested the co-addition process in \citetalias{schl11}
and obtained good agreement between globular cluster metallicities
produced by co-addition and their known metallicities from high-resolution
spectroscopy.

We also determine an equivalent $ugriz$ photometric measurement for our
co-add spectra by computing a weighted average of the $ugriz$ photometric
measurement of the individual stars in each bootstrap co-add, using the
mean S/N between 3950 \AA~and 6000 \AA~as the weight.

We use the SEGUE Stellar Parameter Pipeline
\citep[SSPP;][]{lee08a,lee08b,lee11,all08,smo11} to determine [Fe/H]
and [$\alpha$/Fe] from the co-added ``average" spectrum.  The SSPP uses
Sloan spectroscopy and $ugriz$ photometry to infer the stellar atmosphere
parameters ($T_{\mathrm{eff}}$, $\log{g}$, [Fe/H], and [$\alpha$/Fe])
of stars observed in the course of SDSS and SEGUE.  The SSPP implements
a multimethod algorithm in which many different techniques are used to
compute the stellar parameters.  The SSPP then averages the result of each
method known to be valid in a given color and S/N range to determine the
final $T_{\mathrm{eff}}$, $\log{g}$, [Fe/H], and [$\alpha$/Fe] reported
for all stars observed in the SDSS and SEGUE surveys.

\clearpage

\clearpage

\clearpage

\clearpage

\clearpage

\clearpage

\clearpage

\clearpage

\clearpage
\LongTables 
\begin{deluxetable}{cccccccccccccccccc}
\tablecaption{Spatial and Radial Distribution of [Fe/H] in the Smooth Component of the Halo\label{tbl-1}}
\tablewidth{0pt}
\tablehead{\colhead{RA} & \colhead{Dec} & \colhead{l} &
\colhead{b} & \colhead{bplate} & \colhead{fplate} &
\colhead{[Fe/H]$_{\mathrm{in}}$}\tablenotemark{a} & \colhead{err} &
\colhead{[Fe/H]$_{\mathrm{out}}$}\tablenotemark{a} & \colhead{err} &
\colhead{[Fe/H]$_{\mathrm{in}}$}\tablenotemark{b} & \colhead{err} &
\colhead{[Fe/H]$_{\mathrm{out}}$}\tablenotemark{b} & \colhead{err} &
\colhead{[Fe/H]$_{\mathrm{in}}$}\tablenotemark{c} & \colhead{err} &
\colhead{[Fe/H]$_{\mathrm{out}}$}\tablenotemark{c} & \colhead{err}\\
\colhead{(deg)} & \colhead{(deg)} & \colhead{(deg)} &
\colhead{(deg)} & \colhead{} & \colhead{} & \colhead{} & \colhead{} &
\colhead{} & \colhead{} & \colhead{} & \colhead{} &
\colhead{} & \colhead{} & \colhead{} & \colhead{} &
\colhead{} & \colhead{}}
\startdata
207.22 &  18.62 &   3.16 &  74.29 & 2905   & 2930   & -2.05  &  0.14  & -1.72  &  0.20  & -1.95  &  0.15  & -1.54  &  0.09  & -1.92  &  0.16  & -1.18  &  0.13  \\
229.44 &   7.18 &   9.84 &  50.00 & 2724   & 2739   & -1.76  &  0.13  & -1.97  &  0.15  & -1.90  &  0.15  &    $\cdots$  &    $\cdots$  & -1.83  &  0.16  &    $\cdots$  &    $\cdots$  \\
243.77 &  16.67 &  31.37 &  41.94 & 2177   & 2188   & -1.48  &  0.10  & -1.59  &  0.17  & -1.54  &  0.09  &    $\cdots$  &    $\cdots$  & -1.55  &  0.08  &    $\cdots$  &    $\cdots$  \\
238.51 &  26.52 &  42.88 &  49.49 & 2459   & 2474   & -2.03  &  0.16  & -1.56  &  0.10  & -2.06  &  0.16  & -1.21  &  0.13  & -1.85  &  0.18  &    $\cdots$  &    $\cdots$  \\
253.12 &  23.95 &  43.95 &  36.06 & 2180   & 2191   & -1.84  &  0.14  & -1.49  &  0.10  & -1.69  &  0.11  &    $\cdots$  &    $\cdots$  & -1.69  &  0.11  &    $\cdots$  &    $\cdots$  \\
320.56 &  -7.18 &  44.84 & -36.65 & 2305   & 2320   & -1.54  &  0.09  & -1.60  &  0.20  & -1.55  &  0.09  &    $\cdots$  &    $\cdots$  & -1.44  &  0.10  &    $\cdots$  &    $\cdots$  \\
311.00 &   0.00 &  46.64 & -24.82 & 1908   & 1909   & -1.59  &  0.17  & -0.71  &  0.08  & -1.49  &  0.10  &    $\cdots$  &    $\cdots$  & -1.49  &  0.10  &    $\cdots$  &    $\cdots$  \\
261.18 &  27.01 &  50.00 &  30.00 & 2182   & 2193   & -1.87  &  0.14  & -0.87  &  0.14  & -1.58  &  0.14  &    $\cdots$  &    $\cdots$  & -1.58  &  0.13  &    $\cdots$  &    $\cdots$  \\
\enddata
\tablenotetext{a}{In/out split at $d \sim 12.1$ kpc}
\tablenotetext{b}{In/out split at $d \sim 14.1$ kpc}
\tablenotetext{c}{In/out split at $d \sim 16.8$ kpc}
\end{deluxetable}
\clearpage 

\clearpage
\LongTables 
\begin{deluxetable}{cccccccccccccccccc}
\tablecaption{Spatial and Radial Distribution of [$\alpha$/Fe] in the Smooth Component of the Halo\label{tbl-2}}
\tablewidth{0pt}
\tablehead{\colhead{RA} & \colhead{Dec} & \colhead{l} &
\colhead{b} & \colhead{bplate} & \colhead{fplate} &
\colhead{[$\alpha$/Fe]$_{\mathrm{in}}$}\tablenotemark{a} & \colhead{err} &
\colhead{[$\alpha$/Fe]$_{\mathrm{out}}$}\tablenotemark{a} & \colhead{err} &
\colhead{[$\alpha$/Fe]$_{\mathrm{in}}$}\tablenotemark{b} & \colhead{err} &
\colhead{[$\alpha$/Fe]$_{\mathrm{out}}$}\tablenotemark{b} & \colhead{err} &
\colhead{[$\alpha$/Fe]$_{\mathrm{in}}$}\tablenotemark{c} & \colhead{err} &
\colhead{[$\alpha$/Fe]$_{\mathrm{out}}$}\tablenotemark{c} & \colhead{err}\\
\colhead{(deg)} & \colhead{(deg)} & \colhead{(deg)} &
\colhead{(deg)} & \colhead{} & \colhead{} & \colhead{} & \colhead{} &
\colhead{} & \colhead{} & \colhead{} & \colhead{} &
\colhead{} & \colhead{} & \colhead{} & \colhead{} &
\colhead{} & \colhead{}}
\startdata
207.22 &  18.62 &   3.16 &  74.29 & 2905   & 2930   &  0.37  &  0.07  &  0.44  &  0.12  &  0.42  &  0.06  &  0.43  &  0.22  &  0.43  &  0.06  &  0.43  &  0.22  \\
229.44 &   7.18 &   9.84 &  50.00 & 2724   & 2739   &  0.26  &  0.16  &  0.45  &  0.14  &  0.22  &  0.14  & $\cdots$  & $\cdots$  &  0.43  &  0.14  & $\cdots$  & $\cdots$  \\
243.77 &  16.67 &  31.37 &  41.94 & 2177   & 2188   &  0.45  &  0.22  &  0.48  &  0.17  &  0.44  &  0.22  & $\cdots$  & $\cdots$  &  0.47  &  0.21  & $\cdots$  & $\cdots$  \\
238.51 &  26.52 &  42.88 &  49.49 & 2459   & 2474   &  0.29  &  0.12  &  0.21  &  0.21  &  0.37  &  0.11  &  0.12  &  0.22  &  0.21  &  0.11  & $\cdots$  & $\cdots$  \\
253.12 &  23.95 &  43.95 &  36.06 & 2180   & 2191   &  0.29  &  0.16  &  0.44  &  0.22  &  0.32  &  0.15  & $\cdots$  & $\cdots$  &  0.32  &  0.15  & $\cdots$  & $\cdots$  \\
320.56 &  -7.18 &  44.84 & -36.65 & 2305   & 2320   &  0.18  &  0.22  &  0.13  &  0.15  &  0.22  &  0.22  & $\cdots$  & $\cdots$  &  0.11  &  0.22  & $\cdots$  & $\cdots$  \\
311.00 &   0.00 &  46.64 & -24.82 & 1908   & 1909   &  0.46  &  0.17  &  0.17  &  0.13  &  0.44  &  0.22  & $\cdots$  & $\cdots$  &  0.44  &  0.22  & $\cdots$  & $\cdots$  \\
261.18 &  27.01 &  50.00 &  30.00 & 2182   & 2193   &  0.39  &  0.15  &  0.04  &  0.18  &  0.31  &  0.18  & $\cdots$  & $\cdots$  &  0.30  &  0.18  & $\cdots$  & $\cdots$  \\
\enddata
\tablenotetext{a}{In/out split at $d \sim 12.1$ kpc}
\tablenotetext{b}{In/out split at $d \sim 14.1$ kpc}
\tablenotetext{c}{In/out split at $d \sim 16.8$ kpc}
\end{deluxetable}
\clearpage 

\clearpage
\begin{deluxetable}{ccccc}
\tablecaption{Mean [Fe/H] in the Smooth Component of the Halo\label{tbl-3}}
\tablewidth{0pt}
\tablehead{\colhead{Volume} & \colhead{Mean [Fe/H]} & \colhead{err} &
\colhead{Mean [Fe/H] Above $|b| = 30^{\circ}$} & \colhead{err}}
\startdata
10.0 kpc $\lesssim d \lesssim$ 12.1 kpc & -1.77 & 0.12 & -1.78 & 0.12 \\
10.0 kpc $\lesssim d \lesssim$ 14.1 kpc & -1.78 & 0.12 & -1.79 & 0.11 \\
10.0 kpc $\lesssim d \lesssim$ 16.8 kpc & -1.72 & 0.12 & -1.77 & 0.12 \\
10.0 kpc $\lesssim d \lesssim$ 17.5 kpc & -1.65 & 0.13 & -1.75 & 0.12 \\
12.1 kpc $\lesssim d \lesssim$ 17.5 kpc & -1.58 & 0.13 & -1.69 & 0.12 \\
14.1 kpc $\lesssim d \lesssim$ 17.5 kpc & -1.55 & 0.13 & -1.63 & 0.12 \\
16.8 kpc $\lesssim d \lesssim$ 17.5 kpc & -1.45 & 0.13 & -1.53 & 0.13
\enddata
\end{deluxetable}

\clearpage
\begin{deluxetable}{ccccc}
\tablecaption{Spatial Autocorrelation in [Fe/H] in the Smooth Component of the Halo\label{tbl-4}}
\tablewidth{0pt}
\tablehead{\colhead{Volume} & \colhead{$I$} & \colhead{E$(I)$} &
\colhead{Var$(I)$} & \colhead{$p$-value}}
\startdata
10.0 kpc $\lesssim d \lesssim$ 12.1 kpc & -0.012 & -0.012 &  0.019 & 1.0e+00 \\
10.0 kpc $\lesssim d \lesssim$ 14.1 kpc &  0.018 & -0.010 &  0.016 & 8.2e-02 \\
10.0 kpc $\lesssim d \lesssim$ 16.8 kpc &  0.049 & -0.009 &  0.015 & 1.2e-04 \\
12.1 kpc $\lesssim d \lesssim$ 17.5 kpc &  0.110 & -0.009 &  0.015 & 4.4e-16 \\
14.1 kpc $\lesssim d \lesssim$ 17.5 kpc &  0.113 & -0.009 &  0.016 & 3.1e-15 \\
16.8 kpc $\lesssim d \lesssim$ 17.5 kpc &  0.077 & -0.010 &  0.018 & 1.3e-06
\enddata
\tablecomments{The five columns are: range in distance from the
Galactic center, Moran's $I$ statistic indicating the degree of spatial
autocorrelation in the data in Table~\ref{tbl-1}, the expected value
of Moran's $I$ statistic under the null hypothesis that there is no
spatial autocorrelation, the variance of the expected value under the
null hypothesis, and the probability that the null hypothesis applies
to the data in Table~\ref{tbl-1}.}
\end{deluxetable}

\clearpage
\begin{deluxetable}{ccccc}
\tablecaption{Spatial Autocorrelation in [Fe/H] Above $|b| = 30^{\circ}$ in the Smooth Component of the Halo\label{tbl-5}}
\tablewidth{0pt}
\tablehead{\colhead{Volume} & \colhead{$I$} & \colhead{E$(I)$} &
\colhead{Var$(I)$} & \colhead{$p$-value}}
\startdata
10.0 kpc $\lesssim d \lesssim$ 12.1 kpc & -0.033 & -0.013 &  0.021 & 3.5e-01 \\
10.0 kpc $\lesssim d \lesssim$ 14.1 kpc & -0.018 & -0.011 &  0.018 & 6.7e-01 \\
10.0 kpc $\lesssim d \lesssim$ 16.8 kpc &  0.026 & -0.010 &  0.017 & 3.5e-02 \\
12.1 kpc $\lesssim d \lesssim$ 17.5 kpc &  0.070 & -0.010 &  0.018 & 4.7e-06 \\
14.1 kpc $\lesssim d \lesssim$ 17.5 kpc &  0.088 & -0.011 &  0.019 & 1.1e-07 \\
16.8 kpc $\lesssim d \lesssim$ 17.5 kpc &  0.067 & -0.012 &  0.022 & 2.1e-04
\enddata
\end{deluxetable}

\clearpage
\begin{deluxetable}{lccccc}
\tablecaption{Spatial Autocorrelation in [Fe/H] in \citet{bul05} Halos\label{tbl-6}}
\tablewidth{0pt}
\tablehead{\colhead{Halo} & \colhead{Volume} & \colhead{$I$} &
\colhead{E$(I)$} & \colhead{Var$(I)$} & \colhead{$p$-value}}
\startdata
1 & 10.0 kpc $< d <$ 12.1 kpc &  0.006 & -0.009 &  0.015 & 3.3e-01 \\
  & 10.0 kpc $< d <$ 14.1 kpc &  0.001 & -0.008 &  0.014 & 5.1e-01 \\
  & 10.0 kpc $< d <$ 16.8 kpc & -0.016 & -0.008 &  0.015 & 5.9e-01 \\
  & 12.1 kpc $< d <$ 20.0 kpc & -0.009 & -0.008 &  0.015 & 9.6e-01 \\
  & 14.1 kpc $< d <$ 20.0 kpc &  0.044 & -0.008 &  0.015 & 4.3e-04 \\
  & 16.8 kpc $< d <$ 20.0 kpc & -0.023 & -0.038 &  0.048 & 7.4e-01 \\
\hline \\[-0.1in]
2 & 10.0 kpc $< d <$ 12.1 kpc &  0.011 & -0.009 &  0.016 & 2.1e-01 \\
\enddata
\end{deluxetable}

\clearpage
\begin{deluxetable}{ccccc}
\tablecaption{Spatial Autocorrelation in [Fe/H] in Via Lactea 2\label{tbl-7}}
\tablewidth{0pt}
\tablehead{\colhead{Volume} & \colhead{$I$} & \colhead{E$(I)$} &
\colhead{Var$(I)$} & \colhead{$p$-value}}
\startdata
10.0 kpc $< d <$ 12.1 kpc & 0.025 & -0.019 &  0.033 & 1.8e-01 \\
10.0 kpc $< d <$ 14.1 kpc & 0.067 & -0.015 &  0.024 & 7.5e-04 \\
10.0 kpc $< d <$ 16.8 kpc & 0.003 & -0.012 &  0.020 & 4.4e-01 \\
12.1 kpc $< d <$ 20.0 kpc & 0.079 & -0.011 &  0.019 & 1.3e-06 \\
14.1 kpc $< d <$ 20.0 kpc & 0.078 & -0.012 &  0.020 & 8.9e-06 \\
16.8 kpc $< d <$ 20.0 kpc & 0.043 & -0.015 &  0.025 & 1.9e-02
\enddata
\end{deluxetable}
\end{document}